\documentclass{article}
\usepackage{amssymb}

\usepackage{graphicx}
\usepackage{amsmath}

\newtheorem{theorem}{Theorem}

\newtheorem{corollary}[theorem]{Corollary}

\newtheorem{definition}[theorem]{Definition}

\newtheorem{lemma}[theorem]{Lemma}

\newtheorem{proposition}[theorem]{Proposition}
\newtheorem{remark}[theorem]{Remark}

\newenvironment{proof}[1][Proof]{\textbf{#1.} }{\ \rule{0.5em}{0.5em}}
\newenvironment{pf}[1][Proof of Proposition]{\textbf{#1} }{\ 
\rule{0.5em}{0.5em}}

\newcommand {\Barr}{\mbox{{\rm Bar}}}
\newcommand {\Gr}{\mbox{{\rm Gr}}}
\newcommand {\Id}{\mbox{{\rm d}}}
\newcommand {\id}{\mbox{{\rm id}}}
\newcommand {\wBar}{\widehat{\mbox{{\rm Bar}}}}
\newcommand{\colim}[1]{\underset{#1}{\mbox{{\rm colim}}}}

\newcommand{\Llim}[1]{\underset{#1}{\lim}}

\newcommand{\Tor}{\mbox{{\rm Tor}}}

\newcommand{\Sym}{\mbox{{\rm Sym}}}
\newcommand {\Ker}{\mbox{{\rm Ker}}}

\newcommand{\dirac}{D\!\!\!\!/}

\begin{document}
\title{Algebraic structure of Yang-Mills theory} 
\author{M. Movshev\\Institut Mittag-Leffler \\Djursholm, Sweden\\ \\ A. 
Schwarz\thanks{The work of both  authors was partially supported by NSF 
grant No. DMS 0204927}\\ Department of Mathematics\\ University of 
California \\ Davis, CA, USA}
\date{\today}

\maketitle
\begin{center}
\vskip .2in
{\it To I. M. Gelfand with admiration}
\vskip  .2in
\end{center}
In the present paper we analyze algebraic structures arising in Yang-Mills 
theory. 
The paper should be considered as a part of a project started with 
\cite{MSch2} and devoted to maximally supersymmetric Yang-Mills theories. 
In this paper we collected those of our results which are correct without 
assumption of supersymmetry and used them to give 
rigorous proofs of some 
results of \cite{MSch2}. 
We consider two different algebraic interpretations of Yang-Mills theory - 
in terms of A$_{\infty}$-algebras
 and in terms of representations of Lie algebras (or associative 
algebras).
 We analyse the relations between these two approaches and calculate 
some  Hochschild (co)homology of algebras in question.

\section{Introduction}
Suppose $\mathfrak{ g}$ is a Lie algebra equipped with nondegenerate inner 
product $<.,.>$. We consider Yang Mills field $A$ as $\mathfrak{ 
g}$-valued 
one-form on complex D-dimensional vector space $\mathbf{ V}$   equipped 
with 
symmetric bilinear inner product $(.,.)$.  (All vector spaces in this paper 
are defined over complex numbers.) We are writing this form as 
$A=\sum_{i=1}^DA_idx^i$, where $x^1,\dots,x^D$   is an orthogonal 
coordinate system on $\mathbf{ V}$ which is fixed for the rest of the 
paper.

We assume that the field $A$ interacts with bosonic and fermionic matter 
fields $\phi,\psi$ which are functions on vector space $\mathbf{ V}$ with 
values in $\Phi\otimes \mathfrak{ g},\Pi \mathbf{ S}\otimes \mathfrak{ g}$ 
respectively. The symbol $\Pi$ stands for the change of parity.
In other words matter fields transform according to adjoint representation 
of $\mathfrak{ g}$. 
 The linear space $\Phi$ is equipped with a symmetric inner product 
$(.,.)$, the linear space $\mathbf{ S}$  is equipped with a symmetric 
bilinear map $\Gamma:\Sym^2(\mathbf{ S})\rightarrow \mathbf{ V}$. An 
important example is 10D SUSY Yang-Mills theory where $D=10$, $\mathbf 
{\Phi}=0, \mathbf {V}$ and $\mathbf {S}$ are spaces of vector and spinor 
representations of $SO(10)$ and $\Gamma$ stands for $SO(10)$-intertwiner $\Sym^2\mathbf {S}\rightarrow \mathbf {V}$. 

  We will always consider the action 
functional $S$ as a holomorphic functional on the space of fields; to 
quantize one integrates $exp(-S)$ over a real slice in this space. (For 
example, if $\mathfrak{ g}=\mathfrak{ gl}(n)$ one takes $\mathfrak{ 
u}(n)$-valued 
gauged fields as a real slice in the space of gauge fields). All 
considerations are local; in other words our fields are polynomials 
or power 
series on $\mathbf{ V}$. This means that the action functionals are 
formal expressions (integration over $\mathbf{ V}$ is ill-defined). 
However 
we work with the equations of motion which are well-defined. It is easy 
to get rid of this nuisance and make definitions completely rigorous. 

Choosing once and for all an orthonormal basis in $\Phi$ and some basis 
in $\mathbf{ S}$ we can identify $\phi, \psi$ with $\mathfrak{ g}$-valued 
fields $(\phi_1,\dots,\phi_{d'})$ and $(\psi^{\alpha})$.
The Lagrangian in these bases takes the form:
\begin{equation}\label{E:vht}
L=\frac{1}{4}\sum_{i,j=1}^D<F_{ij},F_{ij}>+\sum_{i=1}^D\sum_{j=1}^{d'}<\nabla_i 
\phi_j,\nabla_i \phi_j>+\sum_{i=1}^D\sum_{\alpha\beta}<\Gamma^i_{\alpha 
\beta} \nabla_i \psi^{\alpha},\psi^{\beta}>-U(\phi,\psi)
\end{equation} 

where $\nabla_i$ stands for covariant derivative built out of $A_i$, 
$F_{ij}=\partial_iA_j-\partial_jA_i+[A_i,A_j]$ denotes gauge field 
strength, $\Gamma^i_{\alpha \beta}$ is the matrix of linear map $\Gamma$ 
in the chosen bases; $U$ is a $\mathfrak{g}$-invariant potential. 
Corresponding action functional $S_{cl}$ is gauge invariant and can be 
extended to solution of BV-master equation in a standard way:
\begin{equation}\label{E:vhtyu}
S=S_{cl}+\int_\mathbf{ V}(\sum_{i=1}^D<\nabla_{i}c,A^{\ast 
i}>+\sum_{\alpha}<[c,\psi^{\alpha}],\psi_{\alpha}^*>+\sum_{j=1}^{d'}<[c,\phi_j],\phi^{*j}>+\frac{1}{2}<[c,c],c^{\ast 
}>)dx^1\dots dx^D
\end{equation} 

Here $c$ stands for Grassmann odd ghost field, $A^{*i}$, 
$\psi_{\alpha}^{*}$, $\phi^{*j}$, $c^*$ are antifields for $A_i$, 
$\psi^{\alpha}$, $\phi_j$, $c$. (The parity of antifields is opposite to 
the parity of fields.)

The BV action functional $S$ determines a vector field $Q$ on the space of 
fields, where $Q$ obeys $Q^2=0$. The space of solutions of equations of 
motion in BV-formalism coincides with zero locus of $Q$. Using $Q$ we 
introduce a structure of L$_{\infty}$-algebra on the space of fields (see 
\cite{AKSZ} or Appendix C in \cite{MSch2}). (Recall that the Taylor 
coefficients of vector field $Q$ obeying $Q^2=0$ at a point belonging to a 
zero locus of $Q$ specify an L$_{\infty}$-algebra. The point in the space 
of fields where all fields vanish belongs to the zero locus of $Q$; we 
construct L$_{\infty}$-algebra using Taylor expansion of $Q$ at this 
point). Equations of motion can be identified with Maurer-Cartan equations 
for L$_{\infty}$-algebra.

The L$_{\infty}$-algebra ${\cal L}$ we constructed depends on the 
choice of Lie algebra $\mathfrak{ g}$ and other data: potential, inner 
products on spaces $\mathbf{ V}$, $\Phi$, bilinear map $\Gamma$ on ${\bold 
S}$. When we need to emphasize such dependence we will do it by 
appropriate subscript: e.g. ${\cal L}_{\mathfrak{ gl}_n}$ shows that the 
Lie algebra $\mathfrak {g}$ is $\mathfrak{ gl}_n$. We will assume that the 
potential $U(\phi,\psi)$ in our algebraic approach is a polynomial (or, 
more generally, formal power series ) of $\phi$ and $\psi$. If $\mathfrak{ 
g}=\mathfrak{ gl}_n$ it has the form $$U(\phi,\psi)=tr(P(\phi,\psi))$$ 
where $P(\phi,\psi)$ is a noncommutative polynomial in matrix fields 
$\phi_1,\dots,\phi_{d'},\psi^{\alpha}$. In this case we construct 
A$_{\infty}$-algebra ${\cal A}$ in a such way that L$_{\infty}$-algebra 
${\cal L}_{\mathfrak{ gl}_n}$ is build in a standard way from A$_{\infty}$-algebra 
${\cal A}\otimes Mat_n$. We can say that working with A$_{\infty}$-algebra 
${\cal A}$ we are working with all algebras ${\cal L}_{\mathfrak{ 
gl}_n}$ at the same time (moreover we can say that we are working with 
gauge theories of all classical gauge groups at the same time.)

We mentioned already that for a $Q$ manifold $X$ (a supermanifold equipped 
with an odd vector field 
$Q$ obeying $Q^2=0$) one can construct an L$_{\infty}$-algebra on the 
vector space $\Pi T^*_{x_0}X$ for every point $x_0\in X$ in zero locus of 
$Q$. In finite dimensional case we can identify L$_{\infty}$-algebra with 
formal $Q$-manifold. On the other hand the algebra of functions on formal 
$Q$-manifold $X$ (= an algebra of formal power series) is a differential 
commutative algebra. This algebra by definition is dual to L$_{\infty}$-algebra 
${\cal L}$.

Similar definitions can be given for A$_{\infty}$-algebras. An algebra of 
functions on formal noncommutative manifold $X$ is defined as topological 
algebra 
of formal noncommutative power series. More precisely if $W$ is a 
$\mathbb{ Z}_2$-graded topological vector space we can consider a tensor 
algebra 
$T(W)=\bigoplus_{n\geq 1} W^{\otimes n}$. This algebra has an additional 
$\mathbb{ Z}$ grading with $n$-th graded component $W^{\otimes n}$ and a 
descending filtration $K^n=\bigoplus_{i\geq n} W^{\otimes i}$. The algebra 
of formal 
power series $\widehat{T(W)}$ is defined as completion of $T(W)$ 
with respect to this filtration. By definition the completion 
$\widehat{T(W)}$ consists of infinite series in generators which become 
finite in projection to $T(W)/K^n$ for every $n$. The elements of 
$\widehat{T(W)}$ are 
infinite sums of monomials formed by elements of a 
basis of $ W$. The algebra $\widehat{T(W)}$ is 
$\mathbb{ Z}_2$-graded; the filtration on $T(W)$ generates a filtration on the 
completion. $\widehat{T(W)}$  can be considered as inverse limit of spaces  
$T(W)/K^n$; we equip  $\widehat{T(W)}$ with the topology of inverse limit 
(the topology on  $T(W)/K^n$ is defined as strongest topology, compatible 
with linear structure.) A 
formal noncommutative $Q$-manifold is by definition a topological algebra 
$\widehat{T(W)}$ equipped with continuous odd differentiation $Q$ obeying 
$Q^2=0$. We say that formal $Q$-manifold $(\widehat{T(W)},Q)$ specifies a 
structure of A$_{\infty}$-coalgebra ${\cal H}$ on the space $\Pi W$. We 
are saying that differential topological algebra $(\widehat{T(W)},Q)$ is 
(bar)-dual to A$_{\infty}$-coalgebra ${\cal H}$. One says also that 
differential algebra  $(\widehat {T(W)},Q)$ is obtained from A$_{\infty}$-coalgebra
 ${\cal H}$ by means of bar-construction ; we 
denote it by $\Barr 
\cal H$.  The homology of $(\widehat{T(W)},Q)$  is called Hochschild 
homology of ${\cal H}$. Notice that in this definition Hochschild homology 
is $\mathbb{ Z}_2$-graded. We obtain also a structure of A$_{\infty}$-algebra 
${\cal A}={\cal H}^*$ 
on $\Pi W^*$. In finite dimensional case the notion of A$_{\infty}$-algebra
 on vector space $V$ is equivalent to the notion of A$_{\infty}$-coalgebra
 on vector space $V^*$. However in infinite-dimensional case it 
is much simpler to use A$_{\infty}$-coalgebras. We will consider the case 
when the  space $W$ is equipped with descending filtration 
$F^n$; then we can extend $F^n$ to filtration of $T(W)$ and  $\widehat{T(W)}$ is defined by means of this filtration . 
(See Appendix , section (\ref{S:mmcsa}) for definitions).

Another way of algebraization of Yang-Mills theory is based on 
consideration of equations of motion (equations are treated as defining 
relations in an associative algebra). We analyse relations between two 
ways of algebraization and study some properties of algebras at hand. In 
particular we calculate some Hochschild homology.

The paper is organized as follows. In Sec. (\ref{S:sdfov}) we 
formulate 
our main results . In Sec. (\ref{S:kcjs}) we give proofs in the case 
of Yang-Mills theory reduced to a point and in Sec. (\ref{S:wfsha},\ref{S:oerdmw})
we consider more general case of Yang-Mills theory 
reduced to any dimension. In Sec. (\ref{S:naiwfc},\ref{F:FFF},\ref{S:nhkshw}) we  make some 
homological calculations  that 
allow us
to apply general results to the case maximally
supersymmetric theories.
 
All proofs in the 
paper are rigorous. However, our exposition in Sec. (\ref{S:kcjs})
is sometimes sketchy; the exposition of more general
results in Sec. (\ref{S:wfsha},\ref{S:oerdmw}) is more formal.

{\bf Notations.}

Denote $<a_1,\dots, a_n>$ a span of vectors $a_1,\dots, a_n$ in some 
linear space.

Denote $\mathbb{ C}<a_1,\dots, a_n>$ a free algebra without a unit  on 
generators $a_1,\dots, a_n$.
If $<a_1,\dots, a_n>=W$ then an alternative notation 
for $\mathbb{ 
C}<a_1,\dots, a_n>$ is 
\begin{equation}\label{E:msjas}
T(W)=\bigoplus_{n\geq 1} W^{\otimes n}
\end{equation}

All algebras in this paper are non- unital algebras , 
unless the opposite is explicitly 
stated.

Any algebra has a canonical filtration $F^n=\{\sum a_1\dots a_n\}$. 

Suppose $A$ is an algebra with a unit and augmentation (i.e homomorphism
 $\varepsilon:A\rightarrow \mathbb{C}$). Denote $I(A)=\Ker \varepsilon$.

Suppose $A$ is an algebra (unit is irrelevant). 
Denote $\underline{A}=A+\mathbb{C}$-an algebra with 
the following multiplicative structure: $(a,\alpha)(b,\beta)=(ab+\alpha b+\beta a,\alpha\beta))$. In this construction we formally adjoint a unit to $A$ equal to $(0,1)$. The algebra $\underline{A}$ has an augmentation $\varepsilon(a,\alpha)=\alpha$ and $I(\underline{A})=A$.

Suppose $\mathbb{Z}_2$-grading of Hochschild homology comes from 
$\mathbb{Z}$-grading. This happens in the case when algebra $A$ has no 
differential, or is $\mathbb{Z}$-graded.
In $\mathbb{Z}$-graded case by our definition zero Hochschild homology  
of algebra without unit is equal to zero (sometime it is called reduced 
homology). Sometimes it will be convenient for us to define 
$H_0(A)=\mathbb{C}$,
 so $H_{\bullet}(A)$ would become an A$_{\infty}$-coalgebra with counit 
and coaugmentation. Such completion will be denoted as $\underline{H_{\bullet}(A)}$. 
It is easy to see that it is equal to standard unreduced Hochschild homology of the unital algebra $\underline{A}$ which is denoted as  $H_{\bullet}(\underline{A},\mathbb{C})$ and defined in \cite{McL}. 
We will use this notation  also in $\mathbb{Z}_2$-graded case.

\subsection{Main results}\label{S:sdfov} 
Let us  reduce the theory to a point, i.e consider the fields that do not 
depend on $x^i,i=1,\dots,D$ (the case of $x$-dependent fields will be 
considered later). 
 The BV-action functional becomes a (super)function and  takes the form:

\begin{equation}\label{E:kdjsdj}
\begin{split}
&S=\frac{1}{4}\sum_{i,j=1}^D<[A_i,A_j],[A_i,A_j]>+\frac{1}{2}\sum_{i=1}^D\sum_{j=1}^{d'}<[A_i 
,\phi_j],[A_i, \phi_j]>+\\
&+\frac{1}{2}\sum_{i,j=1}^D\sum_{\alpha 
\beta}\Gamma^i_{\alpha\beta}<[A_i,\psi^{\alpha}],\psi^{\beta}>-U(\phi,\psi)+\\
&+\sum_{i=1}^D<[A_{i},c], A^{ 
i*}>+\sum_{j=1}^{d'}<[c,\phi_j],\phi^{*j}>+\sum_{\alpha}<[c,\psi^{\alpha}],\psi^*_{\alpha}>+\frac{1}{2}<[c,c],c^{\ast 
}>
\end{split}
\end{equation}

Here $A_i,\phi_i, \psi^{\alpha},c^*$ are elements of $\mathfrak {g}$, and 
$A^{\ast i},\phi^{\ast i}, \psi_{\alpha}^{\ast}$ and $c$ are elements of 
$\Pi\mathfrak {g}$.
The vector field $Q$, corresponding to $S$ is given by the formulas:
\begin{equation}\label{E:yunbr}
\begin{split}
&Q(A_i)=[c,A_i]\\
&Q(\phi_j)=[c,\phi_j]\\
&Q(\psi^{\alpha})=[c,\psi^{\alpha}]\\
&Q(c)=\frac{1}{2}[c,c]\\
&Q(c^*)=\sum_{i=1}^D[A_i,A^{*i}]+\sum_{j=1}^{d'}[\phi_j,\phi^{*j}]+\sum_{\alpha}\{\psi^{\alpha},\psi_{\alpha}^{*}\}+[c,c^*]\\
&Q(A^{*m})=-\sum_{i=1}^D[A_{i},[A_{i},A_{m}]]-\sum_{k=1}^{d'}[\phi_k[\phi_k,A_{m}]]+\frac{1}{2}\sum_{\alpha 
\beta}\Gamma_{\alpha \beta}^m\{\psi^{\alpha},\psi^{\beta}\}-[c,A^{*m}]\\
&Q(\phi^{*j})=-\sum_{i=1}^D[A_i[A_i,\phi_j]]-\frac{\partial U}{\partial 
\phi_j}-[c,\phi^{*j}]\\
&Q(\psi_{\alpha}^{*s})=-\sum_{i=1}^D\sum_{\beta}\Gamma_{\alpha 
\beta}^i[A_i,\psi^{\beta}]-\frac{\partial U}{\partial 
\psi^{\alpha}}-[c,\psi_{\alpha}^{*}]\\
\end{split}
\end{equation}

Let us consider the case $\mathfrak{ g}=\mathfrak{ gl}(n)$. In this case 
all fields are matrix valued functions. In order to pass from L$_{\infty}$ 
to A$_{\infty}$ construction we need to assume that the functions $ 
u_i(x)$ are equal to matrix polynomials.
 
We can construct such an A$_{\infty}$-algebra ${\cal A}_0$ that the 
L$_{\infty}$-algebra ${\cal L}_{\mathfrak{ gl}(n)}$ can be obtained as 
L$_{\infty}$-algebra corresponding to A$_{\infty}$-algebra algebra ${\cal 
A}_0\otimes Mat_n$. The construction is obvious - we consider 
$A_i,\phi_j,c, A^{*i},\phi^{*j},c^*$ in (\ref{E:yunbr}) not as matrices 
but as formal generators. Then $Q$ determines derivation $\hat{Q}$ in 
algebra $\widehat{T(W)}$ of formal power series with respect to free 
generators (we consider $A_i,\phi_j,\psi^*_{\alpha},c^*$ as even elements 
and $A^{*i},\phi^{*j},\psi^{\alpha},c$ as odd ones). The space $W$ can be 
considered as a direct sum of spaces $V,\mathbf {\Phi}, \Pi \mathbf {S}, 
\Pi {\bf C},\Pi \mathbf {V},\Pi \mathbf {\Phi}, \mathbf {S}^*, {\bf C}$.

The derivation $\hat{Q}$ obeys $\hat{Q}^2=0$, hence it specifies a 
structure of A$_{\infty}$-coalgebra on $\Pi W$. The potential $U(\phi, 
\psi)$ is a linear combination of cyclic words in alphabet 
$\phi_{j},\psi^{\alpha}$. In other words $U  (\varphi, \psi)$ is an 
element of ${\rm Cyc}W=T(W)/[T(W),T(W)]$ or, if we allow infinite sums, an 
element of completion of this space. The linear space $[T(W),T(W)]$ is 
spanned by $\mathbb{Z}_2$-graded commutators. Notice, that for every $w\in W^*$ one 
can define derivative $\partial/\partial w:{\rm Cyc}W\rightarrow T(W)$; this map can be 
extended to completions. The derivative $\partial U/\partial \phi _j$ in 
the definition of operator $Q$ should be understood in this way. The 
derivation $\hat{Q}$  specifies not only a structure of A$_{\infty}$-coalgebra
 on $\Pi W$ but also a structure of A$_{\infty}$-algebra on $\Pi 
W^*$. Let us consider for simplicity the case of bosonic theory; writing 
the potential in the form $$U(\phi)=\sum_k c^{j_1, \dots, j_k}\phi_1 \dots 
\phi_k$$ we can represent the operations in the A$_{\infty}$-algebra in 
the following way: 

\begin{equation}
\begin{split}
&m_k(\mathbf{ p}^{j_1},\dots,\mathbf{ 
p}^{j_k})=-c^{j_1,\dots,j_{k+1}}\mathbf{ p}^{*}_{j_{k+1}}\quad(k\geq 2)\\
&m_2(\mathbf{ a}^{{i_1}},\mathbf{ a}_{ {i_2}}^*)=m_2(\mathbf{ 
a}_{i_2}^*,\mathbf{ a}^{{i_1}})=-\bar{\mathbf{ c}}^{\ast 
}\delta^{i_1}_{i_2}\\
&m_2(\mathbf{ p}^{{j_1}},\mathbf{ p}_{ j_2}^*)=m_2(\mathbf{ p}_{ 
{j_2}}^*,\mathbf{ p}^{{j_1}})=-\bar{\mathbf{ c}}^{\ast 
}\delta^{j_1}_{j_2}\\
&m_3(\mathbf{ a}^{{i_1}},\mathbf{ a}^{{i_2}},\mathbf{ 
a}^{{i_3}})=-(\delta^{{i_1}{i_2}}\mathbf{ a}_{ 
{i_3}}^*-2\delta_{{i_1}{i_3}}\mathbf{ a}_{ 
{i_2}}^*+\delta^{{i_2}{i_3}}\mathbf{ a}_{ {i_1}}^*)\\
&m_3(\mathbf{ a}^{{i_1}},\mathbf{ a}^{{i_2}},\mathbf{ 
p}^{j})=-\delta_{{i_1}{i_2}}\mathbf{ p}_{ j}^*\\
&m_3(\mathbf{ a}^{{i_1}},\mathbf{ p}^{j},\mathbf{ 
a}^{{i_2}})=2\delta_{{i_1}{i_2}}\mathbf{ p}_{ j}^*\\
&m_3(\mathbf{ p}^{j},\mathbf{ a}^{{i_1}},\mathbf{ 
a}^{{i_2}})=-\delta_{{i_1}{i_2}}\mathbf{ p}_{ j}^*\\
&m_3(\mathbf{ p}^{{j_1}},\mathbf{ p}^{{j_2}},\mathbf{ 
a}^{i})=-\delta_{{j_1}{j_2}}\mathbf{ a}_{ i}^*\\
&m_3(\mathbf{ p}^{{j_1}},\mathbf{ a}^{i},\mathbf{ 
p}^{{j_2}})=2\delta_{{j_1}{j_2}}\mathbf{ a}_{ i}^*\\
&m_3(\mathbf{ a}^{i},\mathbf{ p}^{{j_1}},\mathbf{ 
p}^{{j_2}})=-\delta_{{j_1}{j_2}}\mathbf{ a}_{ i}^*
\end{split}
\end{equation}
(We are using a basis of $\Pi W^*$ that is dual to the basis of $W$.)
There is an additional set of equations relating $\mathbf{ c}$ with the 
rest of the algebra. They simply assert that $\mathbf{ c}$ is a unit.

The ideal $I(c)\subset {T(W)}$ generated by element $c$ is closed under 
differential $\hat{Q}$. Denote $BV_0$ the quotient differential algebra 
$T(W)/I(c)$. To define a filtration on $BV_0$ we introduce first of all 
grading on $W$ assuming  that $$deg(c)=0,\  deg(A_i)=deg(\phi_i)=2,\  
deg(\psi^{\alpha})=3,\   deg(\psi^*_{\alpha})=5,\ deg(A^{*i})=deg(\phi^{*i})=6,\  deg( c^*)=8.$$
Corresponding multiplicative grading on $T(W)$ in general does not 
descends to $BW_0$, but the decreasing filtration on $T(W)$ generated by 
grading descends to $BV_0$ and is compatible with the differential $Q$ in 
the case when $deg(U) \geq 8$. (The grading on $W$ induces a grading on 
the space ${\rm Cyc} (\Phi +\Pi S)$ of cyclic words and corresponding filtration 
$ F^k$; the notation  $deg(U)\geq 8$ means that  $U\in  F^8$.)1

We always impose the condition  $deg(U)\geq 8$ considering $BV_0$; under 
this condition we can consider the   completion   of $ BV_0$ as filtered 
differential algebra 
 $\widehat{BV_0}$ that can be identified with the quotient algebra 
$\widehat{T(W)}/\widehat{I(c)}$.
Notice that $Q$ is a polynomial vector field hence instead of the 
algebra $\widehat{T(W)}$ of formal power series we can work with tensor 
algebra $T(W)=\bigoplus_{k\geq 0} W^{\otimes k}$, however without 
additional assumptions on the potential $U$ the results of our paper are 
valid only for completed algebra $\widehat{T(W)}$. The differential 
algebra $(\widehat{T(W)},\hat{Q})$ is dual to A$_{\infty}$-algebra ${\cal 
A}$. 
  It will be more convenient for us to work with A$_{\infty}$-coalgebras. 
The motivation is that we would like to avoid dualization in the category 
of infinite dimensional vector spaces as much as possible. However there 
is an involutive duality functor on the category of finite-dimensional or 
graded vector spaces. It implies that a category of finite-dimensional 
A$_{\infty}$-algebras is dual to the category of A$_{\infty}$-coalgebras. 
The same statement is true for category of A${_{\infty}}$-(co)algebras 
equipped with additional grading. There is a topological version of such 
duality which is not auto-equivalence of the appropriate category, rather 
is equivalence between two different categories. 
 
Another approach to algebraization of Yang-Mills theory is based on 
consideration of equations of motion. We will illustrate it in the case of 
Yang-Mills theory, reduced to a point.  We consider its equations of 
motion 

\begin{align}
&\sum_{i=1}^D[A_{i},[A_{i},A_{m}]]+\sum_{k=1}^{d'}[\phi_k[\phi_k,A_{m}]]- \notag \\
&-\frac{1}{2}\sum_{\alpha\beta}\Gamma_{\alpha\beta}^m\{\psi^{\alpha},\psi^{\beta}\}=0 \quad m=1,\dots,D \label{E:relssfs1}\\
&\sum_{i=1}^D[A_i[A_i.\phi_j]]+\frac{\partial U}{\partial \phi_j}=0 \quad j=1,\dots,d'\label{E:relssfs2}\\
&\sum_{\beta}\sum_{i=1}^D\Gamma_{\alpha\beta}^i[A_i,\psi^{\beta}]+\frac{\partial U}{\partial \psi^{\alpha}}=0 \label{E:relssfs3}
\end{align}

as defining relations of 
associative 
algebra with generators $A_i,\phi_j, \psi^{\alpha}$. This algebra will be 
denoted by $YM_0$.
(The algebras $\cal A$ and $YM_0$ depend on the choice of potential $U$, 
hence more accurate notations 
should be ${\cal A}_0^U$ and $YM_0^U$). One can say that $YM_0$ is a 
quotient of tensor algebra $T(W_1)$ with generators  $A_i,\phi_j, 
\psi^{\alpha}$ with respect to some ideal. The grading on $W_1=\mathbf{V}+\Phi +\Pi 
\mathbf{S}$ generates grading on $YM_0$ if 
 $deg(U) =8$. It generates a decreasing filtration compatible with algebra 
structure on $YM_0$ if  $deg(U) 
\geq 8$; in this case we can introduce an algebra structure on the 
completion $\widehat {YM_0}$. Graded algebra associated with the filtered 
algebra  $YM_0$ will be denoted by $YM_0'$; it can be described as the 
algebra $YM_0$ corresponding to component of $U$ having degree $8$.

\begin{theorem}\label{T:fff}

If the potential $U$ has degree $\geq 8$, then the differential 
algebra $(\widehat{BV_0},\hat{Q})$ is quasiisomorphic to the algebra 
$\widehat{YM_0}$.
If the potential $U$ has degree 8 we can say also that
the differential algebra $(BV_0, Q)$ is quasiisomorphic to the algebra 
$YM_0$.
\end{theorem}

\begin{proof}

See section (\ref{S:kcjs}) for the proof.
\end{proof}

The above constructions can be included into the following general scheme.

Let us consider $\mathbb{Z}_2$-graded vector spaces $W_1=V$ with basis 
$e_1,\dots,e_n$, $W_2=\Pi V^*$ with dual basis $e^{*1},\dots, 
e^{*n}$, and one-dimensional spaces $W_0$ with odd generator $c,\  W_3=\Pi 
W_0^*$ with even generator $c^*$. Take $L \in {\rm Cyc } (W_1)$.

Define a differential $Q$ on a free algebra $T(W)$   by the rule:
\begin{equation}\label{E:kjxsw}
\begin{split}
& Q(e_i)=[c,e_i]  \\
& Q(\bar{e}^i)=\frac{\partial L}{\partial {e}^i}+[c,e^{*i}]\\
& Q(c)=-\frac{1}{2}[c,c] \\
& Q(c^*)=\sum_{i}[e_i,\bar{e}^i]+[c,c^*]
\end{split}
\end{equation}

Denote $W^{red}=W_1+W_2+W_3$.
Denote the algebra $T(W)$ with differential $Q$ defined by formula 
(\ref{E:kjxsw}) as $T(W)^L$.
Denote $I(c)$ an ideal generated by $c$. It is easy to see that it is a 
differential ideal. Denote $BV^L=T(W)^L/I(c)$. 

It is easy to see that $BV_0=BV^L$ for   $L$ defined in (\ref{E:vht}), where 
one should disregard $<>$ signs.(We consider $A_i,\phi_j,\dots$ as  free 
variables and $L$ as a $\mathbb{Z}_2$-graded cyclic word.)
The algebra $YM^L$ is defined as a quotient of $T(W_1)$ with respect to 
the ideal generated by $\partial L/\partial e^i$. There exists a natural 
homomorphism of $BV^L$ onto $YM^L$ that sends $e_i \rightarrow e_i$,\  
$e^{*i}\rightarrow 0,\ ,\  \  ,c^* \rightarrow 0$. If 
this homomorphism is a quasiisomorphism we say that $L$ is regular. 
Theorem 1 gives a sufficient condition of regularity.

Theorem 1 can be generalized to unreduced Yang-Mills theory or to theory 
reduced to $d$ dimensions $0 < d \leq D$ . Our consideration will be 
local; this means that for theory with gauge group $\mathfrak{ g}$ reduced 
to $d$ dimensions we consider $\mathfrak{ g}$-valued fields $A_i, 
A^{*i},\phi_j,\phi^{*j},c,c^*$ that are formal functions of the first $d$ 
variables. They span a space $W_d\otimes \mathfrak{g}$, where  $W_d=W\otimes \mathbb{ C}[[x^1,\dots,x^d]]$. 
The space $W_d$ is equipped with filtration $F^s$, which induces filtration
 on $W_d\otimes \mathfrak{g}$ in a trivial way. The group $F^s$ 
consists of all power series with coefficients in $W$ with Taylor 
coefficients vanishing up to degree $s$. The filtration $F^s$ defines a 
topology in a standard way.

 The solutions of the equations of motion in BV-formalism correspond to 
zeros of vector field $Q$  defining a structure of A$_{\infty}$-coalgebra 
on the space $\Pi W_d$; we denote this coalgebra by $bv_d$.
 We can also work with A$_{\infty}$-algebra defined on the space $\Pi W^* 
\otimes \mathbb{ C}[x_1,\dots,x_d]$.

The role of algebra $YM_0$ in the case at hand is played by truncated 
Yang-Mills 
algebra $T_dYM$. We consider a set of differentiations $\partial_k: \underline{YM}_0 
\rightarrow \underline{YM}_0, k=1,\dots,d$. The differentiations are defined by the 
formula: 
\begin{equation}\label{E:jshgsg}
\begin{split}
& \partial_k (A_i) = \delta_{ki} \quad 1 \leq i \leq d \\
& \partial_k(A_i) = 0 \quad i>d \\
& \partial_k(\phi_j)=0\quad \mbox{ for all }j.
\end{split}
\end{equation}

\begin{definition}

We define $\underline{T_dYM}$ as $\bigcap_{k=1}^d Ker \partial_k$. We assume that 
$\partial_k,k=1,\dots,d$ is generic and the matrix $\Gamma$ is not 
degenerate. We define $T_dYM$ in a standard way as $I(\underline{T_dYM})$.
\end{definition}
The precise meaning of assumptions in this definition can be explained in 
the following way. 
\begin{definition}\label{D:jkgjndxcn}
{\rm We say that the set of differentiations $\partial_k,k=1,\dots,d\quad $ is 
generic if the restriction of bilinear form from $\mathbf{ V}$ to 
$T_dYM\cap \mathbf{ V}$ is non- degenerate. If the set of generators 
include fermions then we require that there is a subspace 
$V'\subset\mathbf{ V}$ of codimension one such that $V'\supset T_dYM\cap 
\mathbf{ V}$ and the bilinear form 
$(s_1,s_2)_v\overset{def}{=}[\Gamma(s_1,s_2)\rightarrow \mathbf{ V}/V']$ 
is non- degenerate}.
\end{definition}
\begin{definition}\label{D:kwsznb}
We say that the matrix $\Gamma$ is non- degenerate if there is at least 
one 
generic differentiation from definition (\ref{D:jkgjndxcn}).
\end{definition}

If we do not assume genericity we use an alternative notation $T_{\mu}YM$ 
for $T_dYM$ which will be adopted throughout the main part of the paper.
The algebra $T_dYM$ is filtered and we can define its completion 
$\widehat{T(W_d)}$. The ideal $I(c)\subset \widehat{T(W_d)}$ generated by 
element $c$ is closed under differential $\hat{Q}$. Denote 
$\widehat{BV_d}$ the quotient algebra $\widehat{T(W_d)}/\widehat{I(c)}$. 
We use notation $\widehat{BV_{\mu}}$ for a quotient 
$\widehat{T_{\mu}(W)}/I(c)$ without assumption of genericity of a family 
$\partial_k,k=1,\dots,d$.
\begin{theorem}\label{T:theorem2}
The differential algebra $(\widehat{BV_{d}},\hat{Q})$ is quasiisomorphic 
to $\widehat{T_dYM}$.
\end{theorem}
\begin{proof} 

See proposition \ref{P:d33q}.
\end{proof}

\begin{corollary}

The algebra $\widehat{T_dYM}$ is quasiisomorphic to the dual to coalgebra 
$bv_d$.
\end{corollary}
\begin{proof} 

See proposition (\ref{P:mhdq}) where the A$_{\infty}$-coalgebra $bv_{d}$ 
is denoted as $bv_{\mu}$.
\end{proof}

Let analyze the structure of the algebra $T_dYM$ for $d \geq 1$.

\begin{definition}
Define an algebra $K(q_1,\dots,q_n|p^1,\dots,p^n;\psi^1,\dots,\psi^{n'})$ 
as a quotient algebra $\mathbb{ C}<q_1,\dots, q_n,p^1,\dots, 
p^n,\psi^1,\dots,\psi^{n'}>/I(\omega)$,
where the ideal $I(\omega)$ is generated by an element 
\begin{equation}\label{E:ufbreo}
\omega=\sum_{i=1}^n [q_i,p^i]-\frac{1}{2}\sum_{j=1}^{n'} 
\{\psi^{j},\psi^{j}\}=0.
\end{equation}
\end{definition}

\begin{theorem}\label{T:neghs}
The algebra $\widehat{T_1YM}$ is isomorphic to the algebra $\widehat{K}$.
\end{theorem} 
\begin{theorem}\label{T:lxdnd}
The algebra $\widehat{T_dYM}$ for $d\geq 2$ is isomorphic to the 
completion of a free algebra $T({\cal H}+ {\cal S}+{\cal G})$, where 
${\cal H}$ stands for the space of all polynomial harmonic two-forms on 
$\mathbb{ C}^d$ and ${\cal S}$ stands for the space of harmonic spinors 
and ${\cal G}$ stands for the space of harmonic polynomials on $\mathbb{ 
R}^d$ with values in $\Phi+\mathbb{C}^{D-d}$.
\end{theorem}
\begin{proof} 

The proofs of statements that are more general then Theorems ( 
\ref{T:neghs}, \ref{T:lxdnd}) are given in {\bf Example 1}, {\bf Example 
2}, {\bf Example 4}, Propositions (\ref{P:yghlf}) and (\ref{P:kfiuw}).
\end{proof}

Notice that above theorems have physical interpretation.

Theorems (\ref{T:fff},\ref{T:theorem2})  can be interpreted as a statement that BV 
formalism 
is equivalent to the more traditional approach to the theory of the 
gauge fields. (One can relate this theorem to the calculation of 
BV-homology 
in \cite{barnich}). Theorem (\ref{T:neghs})  is related 
to Hamiltonian formalism in the gauge theory when we neglect the 
dependence of fields on all spatial variables. A solution to the equation 
of motion in such a theory is characterized by a point of a phase space; 
the degeneracy of Lagrangian leads to constraint (\ref{E:ufbreo}) on 
the phase space variables.

Theorems (\ref {T:neghs}) and (\ref{T:lxdnd}) mean that there exists a 
one-to-one correspondence 
between solutions of full Yang-Mills equation of motion and solutions of 
linearized version of this equation.

The phase space dynamics specifies an action of one-dimensional Lie 
algebra
 $\mathfrak{a}$ on algebra $K$. More precisely, we define an action of 
exterior derivation $H$, corresponding to generator of $\mathfrak{a}$ by 
the rule

\begin{align}
&H(q_i)=p^i \label{E:act1}\\
&H(Q_i)=P^i\label{E:act2}
\end{align}
\begin{equation}\label{E:act3}
H(p^i)=-\sum_{i=1}^{D-1}[q_i,[q_i,q_m]]-\sum_{j=1}^{d'}[Q_j,[Q_j,q_m]]+\frac{1}{2}\sum_{\alpha\beta}\Gamma_{\alpha\beta}^m\{\psi^{\alpha},\psi^{\beta}\} 
\end{equation}
\begin{equation}\label{E:act4}
H(P^j)=-\sum_{i=1}^{D-1}[q_i,[q_i,Q_j]]-\frac{\partial U}{\partial Q_j}
\end{equation}
\begin{equation}\label{E:act5}
H(\psi^{\alpha})=-\sum_{i=1}^{D-1} \sum_{ 
\beta}\Gamma_{\alpha\beta}^i[q_i,\psi^{\beta}]-\frac{\partial U}{\partial 
\psi^{\alpha}}
\end{equation}

\begin{definition}\label{D:wddgfh}
Suppose $A$ is an algebra with a unit and $\mathfrak{ g}$ is a Lie algebra which acts 
upon $A$ via derivations. Let 
\begin{equation}
\rho:\mathfrak{ g}\rightarrow Der A
\end{equation}
the homomorphism to the Lie algebra of derivations that corresponds to the action. Let $U(\mathfrak{ g})$ be 
the universal enveloping algebra of $\mathfrak{ g}$. Denote by 
$U(\mathfrak{ g})\ltimes A$ the algebra defined on space $U(\mathfrak{ 
g})\otimes A$ in the following way. It contains $U(\mathfrak{ g})\otimes 
1$ and $1\otimes A$ as subalgebras. For $g\in U(\mathfrak{ g}) $ and $a\in 
A$, $ga=g\otimes a$. If $g$ is a linear generator of $\mathfrak{ g}$ then 
$ga-ag=\rho(g)a$. We call $U(\mathfrak{ g})\ltimes A$ a semi-direct 
product.
\end{definition}

\begin{theorem}\label{T:obms}
Suppose $\Gamma$-matrices used in the definition of $YM_0$ are not 
degenerate in a sense of definition (\ref{D:kwsznb}). Then the algebra 
$\underline{YM}_0$ is isomorphic to a semidirect product 
\begin{equation}\label{E:mkkdjc}
\underline{YM}_0=U(\mathfrak{ a})\ltimes \underline{K}(q_1,\dots, q_{D-1},Q_1,\dots, 
Q_{d'}|p^1,\dots, p^{D-1},P^1,\dots, P^{d'};\psi^{\alpha}).
\end{equation}

 In the above formula $U(\mathfrak{ a})$ is a universal enveloping algebra 
of an abelian one-dimensional Lie algebra $\mathfrak{ a}$ spanned by 
element $H$, which acts on $\widehat{K}$ as an outer differentiation.

The formula (\ref{E:mkkdjc}) remains valid for completed algebras 
$\widehat{\underline{YM}_0}$ and $\widehat{\underline{K}}$ if we replace semidirect product with its 
completion with respect to multiplicative decreasing  filtration $F^n$ 
which coincides with intrinsic filtration on $\widehat{\underline{K}}$ and is 
determined by condition $H\in F^2$.
\end{theorem}

\begin{proof} 

See section (\ref{S:kcjs}) for the proof.
\end{proof}

Using general results about Hochschild homology (the main reference is \cite{McL}, see also Appendix) for algebras with one 
relation and on homology of a cross-product we get the following theorem:

\begin{theorem}\label{T:YM1}
The Hochschild homology 
$\underline{H}_*(K(q_1,\dots,q_n|p^1,\dots,p^n;\psi^1,\dots,\psi^{n'}))$ are 
isomorphic to 
\begin{align}
&\underline{H}_0(K)=\mathbb{C}\\
&\underline{H}_1(K)=<[q_1],\dots,[q_n],[p_1],\dots,[p_n],[\psi^1],\dots,[\psi^{n'}]>\\
&\underline{H}_2(K)=<r>\\
&\underline{H}_{i}(K)=0 \mbox{ for }i\geq 3.
\end{align}
The symbols 
$[q_1],\dots,[q_n],[p_1],\dots,[p_n],[\psi^1],\dots,[\psi^{n'}]$  are in 
one to one correspondence with generators $q_1,\dots, q_n, p_1,\dots, 
p_n,\psi^1,\dots,\psi^{n'}$. There is a nondegenerate even skew-symmetric 
pairing on $\underline{H}_1(K)$ which depends on a choice of a generator $r$ in 
$\underline{H}_2(K)$.
The statement of proposition holds if one replaces algebra $K$ by its 
completion $\widehat{K}$.
\end{theorem}

\begin{proof} 

The section (\ref{S:kcjs}) for the proof.
\end{proof}

\begin{theorem}\label{T:mbdq}
The Hochschild homology 
$\underline{H}_*(\widehat{YM_0})$ are isomorphic to 
\begin{equation}\label{E:kkkhshjs}
\begin{split}
&\underline{H}_0(\widehat{YM_0})=W_0=<c>\\
&\underline{H}_1(\widehat{YM_0})=W_1= \Phi + \Pi \mathbf{S}\\
&\underline{H}_2(\widehat{YM_0})=W_2= \Pi W_1^*\\
&\underline{H}_3(\widehat{YM_0})=W_3=<c^*>. 
\end{split}
\end{equation}
There is a graded commutative duality pairing $$\underline{H}_i(\widehat{YM_0})\otimes 
\underline{H}_{3-i}(\widehat{YM_0})\rightarrow \mathbb{ C}$$ which depends on the 
choice of a generator $c^*$ in $W_3$.

The algebra $YM_0$ has the same Hochschild homology.
\end{theorem}

\begin{proof} 

See the section (\ref{S:kcjs}) for the proof.
\end{proof}

The duality of A$_{\infty}$-algebras is closely related to Koszul 
duality of quadratic algebras (see e.g. \cite{rqwerqe}). 

Let $\mathbf{ S}$ be a spinor representation of $Spin(10)$. Let $${\cal 
S}=\Sym(\mathbf{ S})/\sum_{\alpha \beta 
}\Gamma^i_{\alpha\beta}u^{\alpha}u^{\beta}$$ where $\Gamma^i_{\alpha 
\beta}$ are spinor $\Gamma$-matrices  and 
$u^1,\dots,u^{16}$ is a basis of $\mathbf{ S}$. The algebra ${\cal S}$ can 
be considered as an algebra of polynomial functions on the space of pure 
spinors (spinors in $\mathbf{ S}^*$ satisfying $\sum_{\alpha \beta 
}\Gamma^i_{\alpha\beta}u^{\alpha}u^{\beta}=0$). 
Denote 
\begin{equation}
B_0={\cal S}\otimes \Lambda(\mathbf{ S})
\end{equation}
with linear generators of $\Lambda(\mathbf{ S})$ denoted by 
$\theta^1,\dots,\theta^{16}$. Define a differential on the algebra $B_0$ 
by the rule:
\begin{equation}
Q(\theta^{\alpha})=u^{\alpha}
\end{equation}
We call the differential algebra $(B_0,Q)$ Berkovits algebra. From now on 
till the end of the section we assume that $YM_0$ is build from the 
following data: $D=10$, $d'=0$, $\mathbf{ S}$ is an irreducible spinor 
representation $\mathfrak{ s}_l$ of $Spin(10)$ , $\Gamma_{\alpha\beta}^i$ 
are the $\Gamma$-matrices associated with spinor representation $\mathbf{ 
S}$. (This means that $YM_0$ is obtained from 10D SUSY YM theory reduced 
to
a point.)

 We checked in \cite{MSch2} that $YM_0$ maps to the Koszul dual to 
Berkovits algebra $(B_0,Q)$ (see \cite{MSch2} and references therein about 
Koszul duality). 
In this paper we prove a statement which was formulated in \cite{MSch2} 
without a proof:
\begin{theorem}\label{T:kd}
Koszul dual to the algebra $(B_0,Q)$ is quasiisomorphic to $YM_0$.
\end{theorem}
\begin{proof} 

To prove this fact we should know the homology of Berkovits algebra. 
Heuristic calculation of this homology was given in \cite{TS}. We present 
rigorous calculation in Sec.(\ref{S:nhkshw}).
\end{proof}

\begin{theorem}\label{T:ksdks}
Berkovits algebra $(B_0,Q)$ is quasiisomorphic to the A$_{\infty}$-algebra 
${\cal A}$ obtained from $D=10$ SUSY YM action functional reduced to a 
point.
\end{theorem}
This statement was formulated in \cite{MSch2}. It follows from 
(\ref{T:kd}) and from relation between Koszul duality and bar-duality. 

Let us consider now $d$-dimensional Berkovits algebra $(B_d,Q)$. 
\begin{definition}\label{D:idnsj}
Berkovits algebra $(B_d,Q)$ is defined as an algebra of polynomial 
functions of pure spinor $u$, odd spinor 
$\theta=(\theta^1,\dots,\theta^{16})$ and commuting coordinates 
$x^1,\dots,x^{d}$ ($d \leq 10 $), equipped with differential:

$$Q=\sum_{\alpha=1}^{16} u^{\alpha}\frac{\partial}{\partial 
\theta^{\alpha}}+\sum_{\alpha \beta 
=1}^{16}\sum_{i=1}^{d}\Gamma_{\alpha\beta}^iu^{\alpha}\theta^{\beta}\frac{\partial}{\partial 
x^i}$$%.
\end{definition}
The algebra $B_d$ is a quadratic algebra. 
\begin{theorem}\label{T:kdmasol}
The Koszul dual to differential algebra $(B_d,Q)$ is quasiisomorphic to 
truncated Yang-Mills algebra $T_{d}YM$. 

\end{theorem}
\begin{proof} 

See propositions (\ref{P:kdjdgb}) and (\ref{P:qraffcsd}).
 \end{proof}

\section{Proofs}

\subsection{Algebras $\widehat{K}$ and $\widehat{YM_0}$}\label{S:kcjs}

\begin{pf}{\bf \ref{T:obms}}
We need to rewrite relations 
(\ref{E:relssfs1},\ref{E:relssfs2},\ref{E:relssfs3}) in a slightly 
different form. Introduce notation $q_i=A_{i+1}(i\geq 1)$, $ 
Q_i=\phi_i(i\geq 1)$, $p^i=[A_1,A_{i+1}](i\geq 1)$, $ P^j=[A_1,\phi_j]$ 
$(j\geq 1)$ . Commutation with $A_1$ preserves the algebra  generated by 
$q_i,Q_i,p^i,P^j,\psi^{\alpha}$. Denote the operation of commutation with 
$A_1$ by $H$: $[A_1,x]=H(x)$. Then by definition we have 
(\ref{E:act1},\ref{E:act2}). 

 In the new notations relation (\ref{E:relssfs1}) when $m=0$ becomes (\ref 
{E:ufbreo}). 
To prove this we use nondegeneracy of $\Gamma$-matrices. It allows us to 
make $\Gamma^1_{\alpha\beta}=\delta_{\alpha\beta}$ by appropriate choice 
of basis in $\mathbf{ S}$ and in $\bold V$.
 The relations (\ref{E:relssfs1}) when $m>0$ become (\ref{E:act3}), the 
relations (\ref{E:relssfs2}) become (\ref{E:act4}) and the relations 
(\ref{E:relssfs2}) become (\ref{E:act5}).
 We can see that in this representation the algebra ${\underline{YM}_0}$ is a 
semidirect product of two algebras: a universal enveloping algebra 
$U(\mathfrak{ a})$ of an abelian algebra with one generator $H$ and of the  
algebra ${\underline{K}}$. 

The action of the universal enveloping algebra is given exterior 
differentiation $H$ (letter $H$ stands for the Hamiltonian) by the 
formulas 
(\ref{E:act1},\ref{E:act2},\ref{E:ufbreo},\ref{E:act3},\ref{E:act4},\ref{E:act5}).

If $H$ acts on $K$ via formulas 
(\ref{E:act1},\ref{E:act2},\ref{E:ufbreo},\ref{E:act3},\ref{E:act4},\ref{E:act5}), 
then the map $p:{\underline{YM}_0}\rightarrow U(\mathfrak{ a})\ltimes \underline{K}$, defined by 
the formulas $$p(A_1)=H,\quad p(A_{i+1})=q_i\quad 1 \leq i\leq D, 
\quad p(\phi_j)=Q_j\quad 1 \leq j\leq d',\quad p(\psi^{\alpha})=\psi^{\alpha} $$ is 
correctly defined and agrees on filtrations. It is because formulas 
(\ref{E:act1},\ref{E:act2},\ref{E:ufbreo},\ref{E:act3},\ref{E:act4},\ref{E:act5}) 
imply (\ref{E:relssfs1},\ref{E:relssfs2},\ref{E:relssfs3}).
As a result $p$ is continuous isomorphism that can be extended to 
completions. 
\end{pf}

We need to formulate basic theorems how to compute Hochschild homology of 
some  algebras:
\begin{proposition}\label{P:h12}\ 
\\

{\bf a.} Suppose $\mathfrak{g}$ is a Lie algebra. Then $H_1(\mathfrak{g},\mathbb{ C})\cong 
\mathfrak{g}/[\mathfrak{g},\mathfrak{g}]$,where 
$[\mathfrak{g},\mathfrak{g}]$ is an ideal of $\mathfrak{g}$ consisting  of 
elements of the form $[a,b]$ where $a,b\in \mathfrak{g} $.

{\bf b.} Suppose $\mathfrak{ g}=F/R$, where $F$ is a free Lie algebra, and $R$ is 
an ideal of relations. Then $H_2(\mathfrak{ g},\mathbb{ C})\cong R\cap 
[F,F]/[F,R]$.
\end{proposition}
\begin{proof} 

See \cite{McL} for the proof.
\end{proof}

\begin{corollary}\label{C:kgdfgh}
Suppose $\mathfrak{g}$ is a positively graded Lie algebra. 

{\bf a.} Let $V\subset \mathfrak{g} $ be a minimal generating subspace 
. Then a canonical map $V \rightarrow \mathfrak{ g}/[\mathfrak{ 
g},\mathfrak{ g}]\cong H_1(\mathfrak{ g},\mathbb{ C}) $ is an isomorphism.

{\bf b.} Suppose $\mathfrak{g}=F/R$, where $F$ is a free algebra, $R$ is an ideal. Assume also that the minimal linear subspace  of relations $L$ which generates the ideal
 $R$ is a subspace of $[F,F]$.  
Then the canonical map $L \rightarrow R/[F,R]=R\cap [F,F]/[F,R]\cong 
H_2(\mathfrak{g},\mathbb{ C})$ is an isomorphism.
\end{corollary}

\begin{proposition}\label{P:cnndjg}
Let $A$ be an algebra complete with respect to decreasing multiplicative filtration $F^s,s \geq1$ 
 such that $\bigcap F^s=0$
. There is an isomorphism between $H_1(A)$ and a minimal linear 
space $X\subset A$ such that the subalgebra generated by $X$ is dense (in a 
sense of topology generated by filtration) in 
$A$.
\end{proposition}
\begin{definition}
Suppose $A$ is an algebra with a unit,  $M$ is a bimodule. If $i\geq 0$ is a minimal 
number such that Hochschild homology $H_{i+k}(A,M)=0$ for all $M$, $k>0$, then one says that 
homological dimension of $A$ is equal to $i$.
\end{definition}
\begin{proposition}\label{P:ytfbd}\cite{H}
Suppose a positively graded algebra $A$ is a quotient of a free algebra $T$ by ideal 
generated by one element $r$. If $r\neq aba$, where $a,b\in T$ and 
$deg(a)>0$, then homological dimension of $\underline{A}$ is equal to $2$ and $H_1(A)$ is isomorphic to a minimal set of generators and $H_2(A)=<r>$
\end{proposition}
\begin{proposition}\label{P:yger}\cite{McL}
There is an isomorphism $H_1(T(W))=W$. All other homology of the free algebra $T(W)$ are trivial.
\end{proposition}

\begin{proposition}\label{p:jkdfuijs}
Let $A$ be a positively graded algebra and $\widehat{A}$ is its completion with 
respect to filtration associated with grading. Then 
$H_*(\widehat{A})=\widehat{H_*(A)}$, where $\widehat{H_*(A)}$ stands for 
the completion of $H_*(A)$ by means of filtration associated with grading.
\end{proposition}
\begin{proof} 

Obvious.
\end{proof}

\begin{pf}{\bf \ref{T:YM1}}.
We can use proposition \ref{P:ytfbd} for computation of homology groups. The 
algebra has only one relation in degree 2 therefore $H_2(K)=<r>$ is one 
dimensional. There is a comultiplication map 
\begin{equation}
\Delta:H_2\rightarrow H_2\otimes H_0+H_1\otimes H_1+H_0\otimes H_2
\end{equation}
The image of the $r$ in the middle component gives the matrix of the 
pairing. It is nondegenerate and after the inversion one gets:
\begin{equation}
\begin{split}
&[q_i]*[p^j]=-[p^j]*[q_i]=\delta_i^j\\
&[\psi^i]*[\psi^j]=[\psi^j]*[\psi^i]=\delta^{ij}
\end{split}
\end{equation}
 and all other products are equal to zero. The vanishing of $H_{i}(K)=0 
\mbox{ for }i\geq 3$ is a corollary of proposition (\ref{P:ytfbd}).
The completion of $K$ is associated with grading,therefore the statement 
of 
the proposition for the completed algebra follows from proposition 
\ref{p:jkdfuijs}.
\end{pf}

\begin{pf}{\bf \ref{T:mbdq}}

By definition we have  isomorphisms $\underline{H}_*(\widehat{YM}_0)=
H_*(\underline{YM}_0,\mathbb{C})$ 
and $\underline{H}_*(\widehat{K})=H_*(\underline{K},\mathbb{C})$.
The algebra $\widehat{\underline{YM}_0}$ contains a dense semidirect product 
$U(\mathfrak{a})\ltimes \widehat{\underline{K}}$ and algebra $\underline{YM}_0$ is equal to 
semidirect product.

Let $A$ is one of these algebras. Introduce an increasing filtration $G^nA$ 
defined as follows. The algebra $U(\mathfrak{a})=\mathbb{C}[H]$ has a 
grading such that $deg(H)=1$, denote by $G^nU(\mathfrak{a})$ the 
associated increasing filtration. Denote $G^n(A)=G^nU(\mathfrak{a})\otimes 
B$, where $B$ is either $\underline{K}$ or $\widehat{\underline{K}}$.
 This filtration induces a 
filtration of Bar-complex. It leads to a spectral sequence which is 
usually attributed to Serre and Hochschild.

The next computations will be carried out for the case of semidirect 
product $U(\mathfrak{a})\ltimes \underline{K} = \underline{YM}_0$. 

The $E^2$ term of the spectral sequence is $E^2_{ij}=H_i(U(\mathfrak{ a}), 
\underline{H}_j(K))=H_i(\mathfrak{ a}, \underline{H}_j (K))$ (here $\underline{H}_i(\mathfrak{ a},\dots)$ 
stands for homology of Lie algebra $\mathfrak{ a}$ with coefficients in 
some module). We have convergence $E^2_{ij}\Rightarrow 
H_{i+j}(U(\mathfrak{a})\ltimes \underline{K},\mathbb{C})$. The homology with trivial coefficients 
of algebra $\underline{K}$ are computed in (\ref{T:YM1}). We have 
$$\underline{H}_1(\widehat{K})=<[q_1],\dots,[q_d],[p^1],\dots,[p^d],[Q_1],\dots,[Q_{d'}],[P^1],\dots,[P^{d'}],[\psi^{\alpha}]]>$$

The action of algebra $\mathfrak{ a}$ or (what is the same) the action of 
its generator $H$ on homology of $K$ is easy to describe. It is trivial on 
$\underline{H}_0(K)$ by obvious reasons. It is trivial on $\underline{H}_2 (K)$ because if one 
apply a differentiation of a free algebra $\mathbb{ C}<q_1,\dots, 
q_{D-1},p^1,\dots, p^{D-1},Q_1,\dots, Q_{d'},P^1,\dots, 
P^{d'},\psi^{\alpha}>$ defined by the formulas 
(\ref{E:act1},\dots, \ref{E:act5}) to a 
LHS of equation (\ref{E:ufbreo}) one gets zero. The action of $H$ on 
$[q_i],[Q_j],[\psi^{\alpha}]\in \underline{H}_1(K)$ is zero and 
$H[p^i]=-[q_i]$, 
$H[P^i]=-[Q_i]$ ( It follows from the formulas 
(\ref{E:act1},\dots,\ref{E:act5}) and restriction on degree of potential.

The differential $d_A:C_1(\mathfrak{ 
a},\underline{H}_1(K))=\underline{H}_1(K)\overset{H}{\rightarrow} 
\underline{H}_1(K)=C_0(\mathfrak{ 
a},\underline{H}_1(K))$(Here $C_i$ denotes group of $i$-th chains of the abelian Lie 
algebra $\mathfrak{ a}$). This is the only nontrivial differential in 
$E_1$ term. The spectral sequence degenerates in $E^2$ term because all 
linear spaces where higher differentials can hit are equal to $\{0\}$. 
This implies that the classes 
$[A_1],\dots,[A_D],[\phi_1],\dots,[\phi_{d'}],[\psi^{\alpha}]$ is a basis 
of $H_1(U(\mathfrak{a})\ltimes \underline{K},\mathbb{C})$. We have a nondegenerate 
pairing on $E^2$ between $E_{i,j}^2$ and $E_{1-i,
2-j}^2$. It comes from the diagonal $E_{1,2}^2\rightarrow E_{i,j}^2\otimes 
E_{1-i,2-j}^2$. This diagonal is a tensor product of the diagonal from 
theorem (\ref{T:YM1}) and a diagonal on homology of abelian Lie algebra. 
It indicates that $H_2(U(\mathfrak{a})\ltimes \underline{K},\mathbb{C})$ is dual to 
$H_1(U(\mathfrak{a})\ltimes \underline{K},\mathbb{C})$ and relations (\ref{E:relssfs1}, 
\ref{E:relssfs2}) is a minimal set of relations, 
$H_3(U(\mathfrak{a})\ltimes \underline{K},\mathbb{C})=\mathbb{C}$. 

All homology groups $H_i(U(\mathfrak{a})\ltimes \underline{K})$ $i\geq 4$ 
are equal to zero because all contributors to these groups in the spectral 
sequnce vanish.

This proves the statement for algebra $U(\mathfrak{a})\ltimes \underline{K}$ 
= $\underline{YM}_0$ . We could not apply the proof  directly to $\widehat{\underline{YM}_0}$ 
because it is not a semidirect product. We have the following argument in 
this case:

The algebra $\widehat{\underline{YM}_0}$ is filtered with $\Gr 
\widehat{\underline{YM}_0}=\widehat{\underline{YM}'_0}$. To define $\widehat{\underline{YM}'_0}$ we need to 
take the potential $U$ of $\widehat{\underline{YM}_0}$ and extract its degree 8 
homogeneous part. The algebra $\underline{YM}'_0$ is graded and its completion with 
respect to decreasing completion associated with the grading is equal to 
$\widehat{\underline{YM}'_0}$. By proposition (\ref{P:cnndjg}) the Hochschild 
cohomology of $\widehat{\underline{YM}'_0}$ is equal to completion of homology of 
$\underline{YM}'_0$. The later are finite-dimensional. They coincide with homology of 
$\underline{YM}'_0$. The spectral sequence associated with filtration 
$F^n\widehat{\underline{YM}_0}$ degenerates in $E^1$ term. The proof follows from this 
fact.

\end{pf}

\begin{definition}
{\rm  
Introduce  a multiplicative filtration on $T(W)$ by extending  filtration 
from  generating space $W$.
After completing algebra $T(W)$ we get $\widehat{T(W)}$. Define a 
continuous differential on the algebra $\widehat{T(W)}$ by the formula 
(\ref{E:yunbr}). 
In assumptions of Theorem (\ref{T:fff})
differential (\ref{E:yunbr}) leaves subalgebra $T(W)\subset 
\widehat{T(W)}$ invariant. Denote the resulting differential algebra by 
$(T(W),Q)$.

}
\end{definition}
Define a map 
\begin{equation}\label{M:uifgmk}
p:\widehat{BV_0}\rightarrow \widehat{YM_0}
\end{equation}
by its values on topological generators:
\begin{equation}
\begin{split}
&p(A_i)=A_i\\
&p(\phi_j)=\phi_j\\
&p(\psi^{\alpha})=\psi^{\alpha}\\
&p(A^{*i})=p(\phi^{*j})=p(\psi^{*}_{\alpha})=p(c^*)=0
\end{split}
\end{equation}
then extend it to the entire algebra using properties of homomorphism and 
continuity.
The maps $p:BV_0\rightarrow YM_0$ and $p:BV'_0\rightarrow YM'_0$ are 
defined by the same formulas.

\begin{pf}{\bf \ref{T:fff}}

We will use Theorems \ref{T:mbdq}
 and their corollaries.  Hochshild homology of $\widehat{YM}_0$ and of $YM_0$ were calculated in Theorem (\ref{T:mbdq}).  To calculate homology of differential algebra $\widehat{BV}_0$ we start with $\widehat{BV}_0$ considered as an algebra without differential. 

It is easy to check that this is a completed free algebra with free 
topological generators: 

$$A_{i},\phi_{i},\psi^{\alpha},A^{*i},\phi^{*i},\psi^{*}_{\alpha},c^*.$$  

It follows from Propositions (\ref{P:yger},\ref{p:jkdfuijs})
 that $H_{1}(\widehat{BV}_0)=W_{1}\oplus W_{1}^{*}\oplus \mathbb{C}$ is the space spanned by free topological generators, $H_{i}(\widehat{BV}_0)=0$  for $i\neq 1$.

Hochshild homology of differential algebra $(\widehat{BV}_0,Q)$ is not $\mathbb{Z}$-graded, but it is $\mathbb{Z}_{2}$-graded. The calculation of completed Hochshild homology of $(\widehat{BV}_0,Q)$ is based on the  general lemma (\ref{L;sssfg}).

In our case $H$ is the coalgebra that corresponds to the differential $Q$ on $\widehat{BV}_0$. This coalgebra  is filtered (in a sence described in Appendix) with $F^1(H)=H, \quad F^2(H)=0$.
The A$_{\infty}$-coalgebra $H$ has differential equal to zero. We obtain that as $\mathbb{Z}_2$-graded vector space Hochshild homology of $(\widehat{BV}_0,Q)$ is isomorphic to Hochshild homology of $\widehat{BV}_0$. 

More precisely, this homology is spanned by the following cocycles, where map ${\iota}$ is identification of generators of $BV_0$ and the homology classes in $\Barr (BV_0,Q)$.
\begin{align}
&{\iota}(A^{*m})=-A^{*m}+\sum_{i=1}^D \left( 
A_{i}|[A_{i},A_{m}]-[A_{i},A_{m}]| A_{i}\right)+\notag\\
&+\sum_{k=1}^{d'}\left(\phi_k|[\phi_k,A_{m}]-[\phi_k,A_{m}]|\phi_k 
\right)+\frac{1}{2}\sum_{\alpha \beta}\Gamma_{\alpha 
\beta}^m\psi^{\alpha}|\psi^{\beta}\label{F:edsf}\\
&{\iota}(\phi^{*j})=-\phi^{*j}+\sum_{i=1}^D\left(A_i|[A_i.\phi_j]-[A_i.\phi_j]|A_i\right)+\widetilde{\frac{\partial 
U}{\partial \phi_j}}\label{E:ytbvsa}\\
&{\iota}(\psi_{\alpha}^{*})=-\psi_{\alpha}^{*s}+\sum_{i=1}^D\sum_{ 
\beta}\left(\Gamma_{\alpha 
\beta}^iA_i|\psi^{\beta}-\psi^{\beta}|A_i\right)+\widetilde{\frac{\partial 
U}{\partial \psi^{\alpha}}}\label{E:gfsfdh}\\
&{\iota}(c^*)=\sum_{i}\left(A_i|(A^{*i}-{\iota}(A^{*i}))-(A^{*i}-{\iota}(A^{*i}))|A_i\right)+ 
\notag \\
&+\sum_{j=1}^{d'}\left(\phi_j|(\phi^{*j}-{\iota}(\phi^{*j}))-(\phi^{*j}-{\iota}(\phi^{*j}))|\phi_j\right)+ 
\notag \\
&+\sum_{ 
\alpha}\left(\psi^{\alpha}|(\psi_{\alpha}^{*}-{\iota}(\psi_{\alpha}^{*}))+(\psi_{\alpha}^{*}-{\iota}(\psi_{\alpha}^{*}))|\psi^{\alpha}\right) 
\label{E:hgfdkj}
\end{align}

The map $\iota$ is the identity map on $A_i,\phi_j,\psi^{\alpha}$.
We need to explain what $\widetilde{}$ means in the formulas 
(\ref{E:gfsfdh}) and (\ref{E:ytbvsa}).
A tensor algebra $T(V)$ generated by linear space $V$ can be considered as 
a free product of algebra $V$ with zero multiplication and an algebra 
spanned by $\otimes$ symbol (the multiplication in such algebra is 
trivial). Consider an algebra spanned by bar symbol $|$ with zero 
multiplication. Then $\Barr  T(V)$ is a subspace of a 
free product 
$V\circ <\otimes> \circ <|>$, where $\circ$ denotes a free product. Define 
a derivation on such algebra by the rule $d(v)=0,v \in W$, 
$d_|(\otimes)=|,d_|(|)=0$. The partial derivatives of $U$ in formulas 
(\ref{E:ytbvsa}, \ref{E:gfsfdh}) are elements of a tensor algebra 
$T(\phi_1,\dots,\phi_{d'},\psi^{\alpha})\subset \Barr 
T(\phi_1,\dots,\phi_{d'},\psi^{\alpha})$. Then $\widetilde{\frac{\partial 
U}{\partial \phi_j}}=d_|\frac{\partial U}{\partial \phi_j}$ and 
$\widetilde{\frac{\partial U}{\partial \psi^{\alpha}}}=d_|\frac{\partial 
U}{\partial \psi^{\alpha}}$, where $d_|$ was extended by continuity to 
$\wBar \widehat{T(V)}$.

Simple direct calculation using formulas (\ref{F:edsf},\dots,\ref{E:hgfdkj}) shows that the homomorphism $(\widehat{BV}_0,Q)\rightarrow \widehat{YM}_0$ induces an isomorphism on Hochshild homology.

\end{pf}
\begin{definition}\label{D:mkcnas}
Denote a minimal model (see \cite{KontsSoi} for definition) for coalgebra $\wBar(\widehat{BV_0},d)$ by 
$\widehat{bv}_0$. The linear space of $\widehat{bv}_0$ coinsides with 
$W_1+W_2+W_3$ and the structure maps coincide with (\ref{E:yunbr})(the 
variable $c$ set to zero). The coalgebras $bv_0$ and $bv'_0$ are 
noncomplete and graded version of coalgebra $\widehat{bv}_0$.
\end{definition}
\begin{proposition}\label{E:kgda}
The theorem (\ref{T:fff}) together with lemma (\ref{L;sssfg}) can 
be rephrased as: A$_{\infty}$-coalgebra $bv_0$ is 
bar-dual to algebra 
$\widehat{YM_0}$.
\end{proposition}

\subsection{Truncated Yang-Mills algebra}\label{S:wfsha}

We need to describe notations adopted in this section.

Define a bigraded vector space $W=\bigoplus_{i=0}^3\bigoplus_{j=0}^8 W_i^j$
\begin{equation}\label{D:uhng1}
\begin{array}{ccc}
W^0_0=<c>&W^2_1=\mathbf{V}+\Phi&W^3_1= \mathbf{S}\\
W^8_3=<c^*>&W^6_2=\mathbf{V}^*+\Phi^*&W^5_2=\mathbf{S}^*.
\end{array}
\end{equation}
We define 
\begin{equation}\label{D:kukf}
W_i=\bigoplus_{j=0}^8 W_i^j \quad W^j=\bigoplus_{i=0}^3 W_i^j.
\end{equation}
We refer to 
\begin{equation}\label{E:gttgds}
\begin{split}
&W=\bigoplus_{i=0}^3W_i\mbox{as to homological and }
W=\bigoplus_{j=0}^8W^j\mbox{as to  additional }
\end{split}
\end{equation}
gradings.
We also have a filtration.
\begin{equation}\label{E:ssr}
F^n(W)_i=\bigoplus_{j\geq k}^8 W_i^j
\end{equation}
The algebra $\widehat{YM_0}$ is filtered by multiplicative filtration $F^n(\widehat{YM_0})$. 
On generating space $W_1$ it is given by the formula (\ref{E:ssr}), which determines it uniquely.
 This filtration was alluded in the end of the proof of proposition  
(\ref{T:mbdq}).

  A continuous differentiation $\partial_i$ of 
$\underline{\widehat{YM_0}}$ is defined by the formulas:

\begin{align}
& \partial_i(A_j)= \delta_{ij} \label{E:pfvx1}\\
& \partial_i(\phi_k)=\partial_i(\psi^{\alpha})=0 
\end{align}
We use $\delta_{ij}$ for Kronecker $\delta$ symbol.
All such differentiations span $D$-dimensional vector space $ \mathbf{ 
V}^*$. They can be arranged into one differentiation 
$\partial:\underline{\widehat{YM_0}} \rightarrow  \mathbf{ V}^*\otimes 
\underline{\widehat{YM_0}}$, 
$\partial(a)=\partial_1(a),\dots,\partial_d(a)$. A choice of projection 
\begin{equation}\label{E:igfd}
\mu:  \mathbf{ V}^* \rightarrow V\rightarrow 0
\end{equation}
 on space $V$ specifies a differentiation $\partial_{\mu}$ of $YM_0$ with 
values in a bimodule $V\otimes \underline{\widehat{YM_0}}$.

 Denote $$\underline{\widehat{TYM_{\mu}}}=\Ker \partial_{\mu}\quad \mbox{{\rm and }}\quad  
\widehat{TYM_{\mu}}=I(\underline{\widehat{TYM_{\mu}}}).$$ The algebra 
$\widehat{TYM_{\mu}}$ is filtered by $\widehat{TYM_{\mu}}\cap F^n$ where 
$F^n=F^n(\widehat{YM_0})$ is a filtration on $\widehat{YM_0}$.

Similar constructions hold for algebras $YM_0$ and $YM'_0$. As a result we 
can define $T_{\mu}YM$ and $T_{\mu}YM'$.
 
Define an algebra 
$\underline{\widehat{E_{\mu}YM}}=\underline{\widehat{YM_0}}\otimes 
\Lambda^{\bullet}(V)$. The differential is defined by the rule:
\begin{equation}
\begin{split}
&d(a)=\partial_{\mu}(a)\in YM_{0}\otimes V\subset YM_0\otimes 
\Lambda^{\bullet}(V) \quad a \in YM_0\\
&d(v)=0 \quad v\in \Lambda^{\bullet}(V).
\end{split}
\end{equation}
The differential can be extended uniquely to the algebra using the Leibniz 
rule.
Denote $$\widehat{E_{\mu}YM}=I(\underline{\widehat{E_{\mu}YM}}).$$

A similar construction works for algebras $YM_0$ and $YM'_0$. We can 
define algebras $E_{\mu}YM$ and $E_{\mu}YM'$. 
Define a multiplicative decreasing  filtration $F^s(\underline{\widehat{E_{\mu}YM}})=F^s$ on $\underline{\widehat{E_{\mu}YM}}$ 
which extends filtration on $F^s(\underline{\widehat{YM_0}})$. It is uniquely 
determined by condition   $V\subset F^1$. It defines a filtration on $\widehat{E_{\mu}YM}$ for which we keep the same notation. A similar filtration exist on 
$E_{\mu}YM$. The algebra $E_{\mu}YM'$ is graded, the grading is a 
multiplicative extension of grading on $YM'_0$, the grading of the space 
$V\subset E_{\mu}YM'$ is equal to one. In case of algebras 
$\widehat{E_{\mu}YM}$, $E_{\mu}YM$, the differential preserves the 
filtration. In case of $E_{\mu}YM'$ the differential preserves the 
grading.

\begin{lemma}\label{L:jhkhg}
Suppose $B$ is an algebra with a unit  generated by elements 
$B_1,\dots,B_n$. Assume that we given $k$ commuting differentiations 
$\partial_s$ $s=1,\dots,k$ $k\leq n$ of the algebra $B$ such that 
$\partial_sB_j=\delta_{sj}$. Then there exists an increasing  filtration 
$G^{i}$ $i=0,1,\dots$, such that 

{\bf a.} $\bigcup_iG^i=B.$

{\bf b.} $G^iG^j\subset G^{i+j}.$

{\bf c.} $B_i\in G^{1}.$%.
\end{lemma}
\begin{proof} 

Define filtration $G^i$ inductively. By definition $G^0=\bigcap_{s=1}^k 
\Ker \partial_s$, then 
\begin{equation}
G^{i+1}=\{x|\partial_s(x)\in G^i \mbox{ for all }s, \quad 1 \leq s \leq k 
\}
\end{equation}
The property {\bf b} follows from the Leibniz rule. By definition $B_i\in 
G^{1}$ hence {\bf c} follows. Since $B_i$ are generators {\bf b} and {\bf 
c} imply {\bf a}.
\end{proof}

Consider an algebra $\Gr_G(B)=\bigoplus^{\infty}_{i=0} G^{i+1}B/G^{i}B$.  
 Denote the image of elements $B_1,\dots, B_k$ in $G^{1}/G^0$ by 
$\hat{B}_1,\dots,\hat{B}_k$.
\begin{lemma} 

The algebra $\Gr_G(B)$ has the following properties:

    {\bf a.}The elements $\hat{B}_1,\dots,\hat{B}_k$ commute in $\Gr_G(B)$.

{\bf b.} The elements $\hat{B}_1,\dots,\hat{B}_k$ commute with $G^{0}B$.

{\bf c.} The subalgebra of $\Gr_G(B)$ generated by $G^{0}B$ and 
$\hat{B}_1,\dots,\hat{B}_k$ is isomorphic to $\mathbb{ 
C}[\hat{B}_1,\dots,\hat{B}_k]\otimes G^{0}B$.

{\bf d.} Elements $\hat{B}_1,\dots,\hat{B}_k$ and $G^{0}B$ generate 
$\Gr_G(B)$.

{\bf f.} We have an isomorphism $\Gr_G(B)=\mathbb{ 
C}[\hat{B}_1,\dots,\hat{B}_k]\otimes G^{0}B$.
\end{lemma}
\begin{proof}\ 

{\bf a.} $\partial_s[B_i,B_j]=[\delta_{si},B_j]+[B_i,\delta_{sj}]=0$, 
therefore $[B_i,B_j]\in G^0$, hence $[\hat{B}_i,\hat{B}_j]=0$.

{\bf b.} Similarly $\partial_s[B_i,m]=[\delta_{si},m]=0$ for $m \in 
G^0\subset \Ker \partial_s$ for $s=1,\dots,k$, Therefore $[B_i,m]\in G^0$ 
and $[\hat{B}_i,\hat{m}]=0$

{\bf c.} Denote the subalgebra generated by $G^{0}B$ and 
$\hat{B}_1,\dots,\hat{B}_k$ by $C$.
There is a surjective map 
\begin{equation}\label{E:jbyuhb}
\mathbb{ C}[\hat{B}_1,\dots,\hat{B}_k]\otimes G^{0}B\rightarrow C
\end{equation}
 and an inclusion $G^0\subset C$. Denote the kernel of the map 
(\ref{E:jbyuhb}) by $I$. Then 
\begin{equation}\label{E:ynbfg}
I\cap G^{0}B=0.
\end{equation} Suppose $0 \neq a=\sum 
a_{i_1,\dots,i_k}\hat{B}^{i_1}_1\dots \hat{B}^{i_k}_k\in I$.  Since $a$ is 
a polynomial in $\hat{B}_1,\dots,\hat{B}_k$, there is $i_1,\dots,i_k$ such 
that there is no $a_{i'_1,\dots,i'_k}\neq0$ with 
$i_1<i'_1,\dots,i_k<i'_k$. It means that an element $\partial^{i_1}_1\dots 
\partial^{i_k}_ka\neq 0$ belongs to $I$ and is independent of 
$\hat{B}_1,\dots,\hat{B}_k$, which contradicts with (\ref{E:ynbfg}).

{\bf d.} We are going to prove the statement by induction on the index $i$ 
in $\Gr^i_G(B)$.
If $i=0$ then there is nothing to prove. Suppose we have an element 
$\hat{a}\in \Gr^{i+1}_G(B)$. Let $a\in G^{i+1}$ its representative in $B$. 
Then $\partial_s a=b_s\in G^{i}$ and by inductive assumption 
$\hat{b}_s=\hat{b}_s(\hat{B}_1,\dots,\hat{B}_k)\in C$. The elements $b_s$ 
satisfy $\partial_i \hat{b}_s=\partial_s \hat{b}_i$. It implies that there 
is $b\in B$ such that $\partial_{s}\hat{b}=\hat{b}_s$. Consider the 
difference $a-b=c$, $\partial_s\hat{c}=0$, hence $\partial_s c \in 
G^{i-1}$ for every $s$, therefore $c\in G^{i}$ and $\hat{a}=\hat{b}$.

\end{proof}

\begin{lemma}\label{L:pjmdfik}
Suppose an algebra $B$ satisfies conditions of lemma (\ref{L:jhkhg}). 
Denote by $V$ a linear space with a basis 
$[\partial_1],\dots,[\partial_k]$.  Define a structure of a complex with 
differential $\sum_{s=1}^k[\partial_s]\partial_s$ on a linear space 
$\Lambda^i(V)\otimes B$. Then the cohomology of this complex are 
concentrated in degree $0$ and isomorphic to $G^0$.
\end{lemma}
\begin{proof} 

Define a filtration on the complex $H_i=B\otimes \Lambda^{i}(V)$ by 
$G^jH_i=G^{i+j}\otimes\Lambda^{i}(V)$.

The adjoint quotients of this filtration are isomorphic to 
\begin{equation}
G^jH_i/G^{j-1}H_i=\Lambda^{i}(V)\otimes \Sym^{(i+j)}(V)\otimes G^0.
\end{equation}
The differential coinsides with de Rham differential. It cohomology is 
isomorphic to $G^0$ in zero degree and $0$ in higher degrees. The spectral 
sequence corresponding to filtration $G^i$ collapses in $E_1$ term and 
converges to cohomology we are looking for.
\end{proof}

\begin{lemma}\label{L:lknb}\ 

{\bf a.}The embeddings $(\underline{TYM_{\mu}},0)\rightarrow 
(\underline{E_{\mu}YM},d)$ and $(\underline{TYM'_{\mu}},0)\rightarrow 
(\underline{E_{\mu}YM'},d)$ are quasiisomorphisms. 

{\bf b.}The map
\begin{equation}\label{E:aasdfv}
(\underline{\widehat{TYM_{\mu}}},0)\rightarrow 
(\underline{\widehat{E_{\mu}YM}},d)
\end{equation} is a filtered quasiisomorphism of algebras (see Appendix for definition).

\end{lemma}
\begin{proof} 

{\bf a.} The algebras $YM_0$ and $YM'_0$ satisfy lemma (\ref{L:pjmdfik}), 
hence the proof follows.

{\bf b.} The morphism (\ref{E:aasdfv}) is compatible with  filtrations 
which exist on its range and domain. It induces a map of spectral 
sequences associated with filtrations. Let us analyse $E^2$-term.

 The algebra $A=\Gr_{F}(\widehat{YM_0})$ is isomorphic to $YM_0'$ where in 
the relations (\ref{E:relssfs2}, \ref{E:relssfs3}) we drop the potential 
$U$. The algebra $YM_0'$ is finitely generated, graded, carries no 
topology. The algebra $A$ satisfies conditions of lemma (\ref{L:pjmdfik}). 
It implies that there is a quasiisomorphism  $(\Gr 
\widehat{TYM}_{\mu},0)\rightarrow (\Lambda(V)\otimes \underline{\widehat{ 
YM_0}},d)$. 

We see that the map (\ref{E:aasdfv}) induces an isomorphism of $E^1$ term 
of corresponding spectral sequences. Since the range and domain are 
complete with respect to filtrations we conclude that the map 
(\ref{E:aasdfv}) is quasiisomorphism. The property that it  induces  
quasiisomorphism of adjoint quotients, it is the same as filtered 
property.
\end{proof}
\begin{corollary}
The map $(\widehat{TYM_{\mu}},0)\rightarrow (\widehat{E_{\mu}YM_0},d)$ is 
a filtered quasiisomorphism. $(TYM_{\mu},0)\rightarrow (E_{\mu}YM,d)$ and 
$(TYM'_{\mu},0)\rightarrow (E_{\mu}YM',d)$ are quasiisomorphisms.
\end{corollary}

\begin{remark}
The algebra $\widehat{E_{\mu}YM}$ is topologically finitely generated. It implies 
that the canonical filtration is comparable with filtration $F^s(\widehat{E_{\mu}YM})$. It 
implies that the algebra $\widehat{E_{\mu}YM_0}$ is complete with respect 
to canonical filtration $I^n_{\widehat{E_{\mu}YM_0}}$ and 
$\widehat{TYM_{\mu}}$ with respect to $\widehat{TYM_{\mu}}\cap 
I^n_{\widehat{E_{\mu}YM_0}}$. 
\end{remark}

\subsection{Construction of $BV_{\mu}$}\label{S:oerdmw}
Let $V$ be a vector space generated by symbols $\partial_i,i=1,\dots,d$, 
$\Sym(V)=\bigoplus_{i=0}^{\infty}\Sym^i(V)$. Denote a decreasing 
filtration of $\Sym(V)$ associated with the grading by $F^n(\Sym(V))$. The 
vector spaces $W_i$ $i=0,\dots,3$ were defined by the formulas 
(\ref{D:kukf}). 
\begin{equation}
\begin{split}
&\overline{W}_i=W_i\otimes \Sym(V)\\
&\overline{W}=\overline{W}_0+\overline{W}_1+\overline{W}_2+\overline{W}_3
\end{split}
\end{equation}
The last direct sum decomposition is called homological grading on 
$\overline{W}$.

Denote two filtrations on $\overline{W}$. The first one is 
\begin{equation}
F^n(\overline{W})=\sum_{i+j\geq n}F^i(W)\otimes F^j(\Sym(V)),
\end{equation}
where $F^i(W)$ was defined in (\ref{E:ssr}).
The second is 
\begin{equation}
\tilde{F}^n(\overline{W})=W\otimes F^n(\Sym(V)).
\end{equation}
We have 
\begin{equation}\label{L:hahahd}
\tilde{F}^n(\overline{W})\subset 
F^n(\overline{W})\subset\tilde{F}^{n-8}(\overline{W})
\end{equation}
with finite dimensional quotients.
Denote the completion of $\overline{W}$ by 
$\widehat{W}=\widehat{W}_0+\widehat{W}_1+\widehat{W}_2+\widehat{W}_3$.

Filtrations $\tilde{F}^n(\overline{W})$ and $F^n(\overline{W})$ define two 
multiplicative filtrations on $T(\overline{W}[1])$, which we denote by 
$\tilde{F}^n(T(\overline{W}[1]))$ and $F^n(T(\overline{W}[1]))$. The 
algebra $T(\overline{W}[1])$ acquires a homological   grading by 
multiplicative extension of homological grading from $\overline{W}$.

Define an additional grading on $\overline{W}$:
\begin{equation}
\overline{W}^k=\bigoplus_{i+j=k}W^i\otimes \Sym^j(V)
\end{equation}
The algebra $T(\overline{W}[1])$ acquires an additional  grading by 
multiplicative extension of additional grading from $\overline{W}$. 

\begin{proposition} 

The completion of $T(\overline{W}[1])$ with respect filtrations 
$\tilde{F}^n(\overline{W})$ and $F^n(\overline{W})$ coincide and we denote 
it by $\widehat{T(\overline{W}[1])}$. Similarly two completions of the 
space of generators $\overline{W}$ coincide.
\end{proposition}
\begin{proof} 

The filtration satisfy the following inclusions:
$\tilde{F}^n(T(\overline{W}[1]))\subset 
F^n(T(\overline{W}[1]))\subset\tilde{F}^{n-k}(T(\overline{W}[1]))$ for 
some finite $k$ with finite-dimensional quotients. This is a simple 
corollary of equation \ref{L:hahahd}. We see that the filtrations are 
commensurable. It is an simple exercise to show  the completions are 
equal.
\end{proof}

 The operators $\partial_i$ act by multiplication on the set of generators 
of the algebra $T(\overline{W}[1])$ (recall that it is a free 
$\Sym(V)$-module). We extend the action of $\partial_i$ on 
$\widehat{T(\overline{W}[1])}$ as a continuous derivation, which we denote 
by the same symbol $\partial_i$. 

There is a linear map $\mu:W_1\rightarrow V$. It is extension by zero from 
$\mathbf{ V}$ to $W_1=\mathbf{ V}+\Pi \mathbf{ S}+\Phi$ of the map $\mu$ 
defined in (\ref{E:igfd}). We used identification $<A_1,\dots,A_D>$ and 
$<A_1,\dots,A_D>^*$ provided by the canonical bilinear form .
The algebra $\widehat{T(\overline{W}[1])}$ admits a continuous  action of 
outer derivation $\nabla_i$ defined by the formula
\begin{equation}\label{E:pnshgrt}
\nabla_ix=\mu(A_i)x+[A_i,x].
\end{equation}
The commutator is defined as:
\begin{equation}
[\nabla_i,\nabla_i]=\mu(A_i)A_j-\mu(A_j)A_i+[A_i,A_j].
\end{equation}
\begin{definition}
{\rm
Define a continuous differential in the algebra 
$\widehat{T(\overline{W}[1])}$ by the formulas (\ref{E:yunbr})
where $deg(U)\geq 8$. Define a  differential on the algebra 
$T(\overline{W}[1])$ by the formula (\ref{E:yunbr}) where $deg(U)\geq 8$ 
and we impose some finitness conditions on the potential.
 We denote such algebra 
$(T(\overline{W}[1]),Q^{\mu})$.
The filtration on algebras $\widehat{T(\overline{W}[1])}$ and 
$T(\overline{W}[1])$ is preserved by the differential.

If one assume that potential $deg(U)=8$ we denote the 
algebra $(T(\overline{W}[1]),Q^{\mu})$ by 
$(T(\overline{W}[1])',Q^{\mu})$. In this algebra the differential has a 
degree zero with respect to the additional grading.

One can extend $Q^{\mu}$ uniquely to the entire set of generators, using 
commutation properties with $\partial_i$.
}
\end{definition}
\begin{proposition} 

The differential $Q^{\mu}$
satisfies $(Q^{\mu})^2=0$. The differential has degree $-1$ with respect 
to 
homological grading.
\end{proposition}
Observe though the elements $\nabla_{i}$ do not belong to the algebra 
$\widehat{T(\overline{W}[1])}$ all RHS expressions in the formulas 
(\ref{E:yunbr}) do.

Define an odd differentiation $\varepsilon$ of the algebra 
$T(\overline{W}[1])$ by the formula $\varepsilon(c)=1$, the value of 
$\varepsilon$ on all other generators is equal to zero. It can be extended 
by continuity to $\widehat{T(\overline{W}[1])}$. 
\begin{proposition} 

The commutator $\{Q^{\mu},\varepsilon\}$ is a differentiation $P$ of the 
algebra $\widehat{T(\overline{W}[1])}$ which on generators is equal to 
identity transformation. The same holds for $T(\overline{W}[1])$.
\end{proposition}
\begin{proof} 

Direct computation.
\end{proof}

 It implies that $P$ is an invertible linear transformation on  
$\widehat{T(\overline{W}[1])}$ and on  $T(\overline{W}[1])$, and 
$H=\varepsilon/P$ is a contracting homotopy.
We are interested in a modification of the algebra 
$(\widehat{T(\overline{W}[1])},Q^{\mu})$, $(T(\overline{W}[1]),Q^{\mu})$, 
$(T(\overline{W}[1])',Q^{\mu})$. Denote $\widehat{I(c)}$ a closure of the 
ideal generated by $c$. A simple observation is that $\widehat{I(c)}$ is a 
differential ideal. The ideal $\widehat{I(c)}$ is not closed under the 
action of $\partial_i$ however.

\begin{definition}
Denote by $\widehat{BV_{\mu}}$ the quotient algebra 
$\widehat{T(\overline{W}[1])}/\widehat{I(c)}$. Similarly define $BV_{\mu}$ 
and $BV'_{\mu}$. The former algebra is filtered the later is graded.
\end{definition}
\begin{remark}
In contrast with algebras $\widehat{BV_{0}}$, $BV_{0}$, $BV'_{0}$ the 
algebras $\widehat{BV_{\mu}}$,$BV_{\mu}$,$BV'_{\mu}$ have nontrivial 
components with negative homological grading.
\end{remark}

\begin{proposition} 

There is a morphism of differential graded algebras 
$(\widehat{BV_{\mu}},Q^{\mu})\rightarrow (\widehat{E_{\mu}YM},d)$
defined by the formulas
\begin{equation}\label{Jnfgfge}
\begin{split}
&p(\partial_ic)=[\partial_i]\\
&p(A_i)=A_i\\
&p(\phi_i)=\phi_i\\
&p(\psi^{\alpha})=\psi^{\alpha}
\end{split}
\end{equation}
which preserves the filtrations (so it is continuous).
The map is zero on the rest of the generators. The formulas 
(\ref{Jnfgfge}) defines a map $(BV_{\mu},Q^{\mu})\rightarrow 
(E_{\mu}YM,d)$ which preserves the filtrations and a map 
$(BV'_{\mu},Q^{\mu})\rightarrow (E_{\mu}YM',d)$ of degree zero.
\end{proposition}

The filtrations $F^n\widehat{T(\overline{W}[1])}$ and 
$\tilde{F}^n\widehat{T(\overline{W}[1])}$ of 
$\widehat{T(\overline{W}[1])}$ induce similarly named  filtrations of 
$\widehat{BV_{\mu}}$ , denoted $F^n\widehat{BV_{\mu}}$ and 
$\tilde{F}^n\widehat{BV_{\mu}}$. Filtration $F^n$ and $\tilde{F}^n$ are 
also defined on $BV_{\mu}$ and $BV'_{\mu}$. In the later case filtration 
$F^n$ coinsides with decreasing filtration associated with the grading.

 Define $\Gr \widehat{BV_{\mu}}=\prod_{n\geq 0} \Gr^n \widehat{BV_{\mu}}$ 
as
\begin{equation}
\Gr^nBV_{\mu}=\tilde{F}^n\widehat{BV_{\mu}}/\tilde{F}^{n+1}\widehat{BV_{\mu}}.
\end{equation} 
\begin{proposition} 

The algebra $\Gr \widehat{BV_{\mu}}$ is a completion of a free algebra  
with the same space of generators as $\widehat{BV_{\mu}}$. The 
differential $Q$ is defined by the  formulas (\ref{E:yunbr}), except in 
the formula (\ref{E:pnshgrt}) one needs to replace 
\begin{equation}\label{E:qsddf}
\mu(A_i)x+[A_i,x] \Rightarrow [A_i,x].
\end{equation}
Similarly $\Gr BV_{\mu}$ and $\Gr BV'_{\mu}$ coincide with $BV_{\mu}$ as 
algebras. In the definition of $Q^{\mu}$ one has to alter $\nabla_i$ 
according to the rule (\ref{E:qsddf}).
\end{proposition}
\begin{proof} 

Obvious.
\end{proof}

Nonreduced bar-complex of a A$_{\infty}$-coalgebra $H$ with a counit 
$\varepsilon$ is by definition bar-complex of $H$ as if it had no counit. 
A simple theorem asserts that in presence of a counit it is always 
contractible.
\ 
\begin{definition}
The algebras $(\widehat{T(\overline{W}[1])},Q^{\mu})$ and $(\Gr 
T(\overline{W}[1]),\Gr Q)$ can be thought of as nonreduced bar-complex of 
A$_{\infty}$-coalgebras with a counit and coaugmentation  which we denote 
by $(\widehat{\overline{W}},Q^{\mu})$ and 
$(\widehat{\overline{W}},\Gr(Q^{\mu}))$, the corresponding coideals are 
denoted as  $\widehat{bv_{\mu}}$ and $\widehat{\Gr bv_{\mu}}$ .
 Similarly we have A$_{\infty}$-coalgebras 
$(\overline{W},Q^{\mu})$,$(\overline{W},\Gr 
Q^{\mu})$;$(\overline{W}',Q^{\mu})$, $(\overline{W}',\Gr Q^{\mu})$ with 
coaugmentation coideals $(bv_{\mu},Q^{\mu})$, $(\Gr bv_{\mu},\Gr Q^{\mu})$ 
and $(bv'_{\mu},Q^{\mu})$, $(\Gr bv'_{\mu},\Gr Q^{\mu})$ respectively.
\end{definition}
There a general construction of tensor product of A$_{\infty}$-coalgebras. 
Its description when one of the tensor factors is an ordinary coalgebra is 
very simple. Suppose $H$, $G$ are two A$_{\infty}$-algebras and $G$ was 
only one nontrivial operation $\Delta_2=\Delta:G\rightarrow G^{\otimes 
2}$. Define a mapping $\nu_n\rightarrow G^{\otimes n}$ by the formula
\begin{equation}
\nu^G_n=(\Delta \otimes \id \otimes \dots \id) \circ \dots \circ \Delta.
\end{equation}
The tensor product of $H$ and $G$ has its underlying vector space equal to 
$H\otimes G$. The operations $\Delta^{H\otimes G}_i$ are defined by the 
formula:
\begin{equation}
\Delta^{H\otimes G}_n(a\otimes b)=T\Delta_n^H(a)\otimes \nu^G_n(b)
\end{equation}
where operator $T$ is a graded permutation which defines isomorphism 
$H^{\otimes n}\otimes G^{\otimes n}\cong (H\otimes G)^{\otimes n}$. We 
need to bring readers attention to the fact that though $\Delta_n^H$ is 
the n-th operation in coalgebra $H$, the map $\nu^G_n$ is not such for 
coalgebra $G$, but rather n-th iteration of the binary operation. As you 
can see this construction is not symmetric.

Observe that on  category of A$_{\infty}$-algebras a similar  operation 
corresponds to extension of the  ring of scalars.

It turns out that the algebras $\widehat{\Gr bv_{\mu}}$, $\Gr bv_{\mu}$, 
$\Gr bv'_{\mu}$ has an alternative description in terms of  finite 
dimensional A$_{\infty}$-coalgebra $\widehat{bv}_0$, $bv_0$, $bv'_0$ 
introduced in definition (\ref{D:mkcnas}).

The Kunneth formula asserts 
\begin{proposition}\label{P:Kund}
There is a quasiisomorphism of algebras $\wBar(H\otimes G)\rightarrow 
\wBar(H)\hat{\otimes}\wBar(G)$.
\end{proposition}
\begin{proof}
See \cite{McL} for the proof in the case of algebras. The coalgebra case is similar.
\end{proof}
\begin{proposition} 

There is an isomorphism of coalgebras $\underline{\widehat{\Gr\ 
bv_{\mu}}}=W\otimes \underline{\widehat{\Sym}(V)}$, $\underline{\Gr\ 
bv_{\mu}}=W\otimes \underline{\Sym(V)}$,  $\underline{\Gr\ 
bv'_{\mu}}=W'\otimes \underline{\Sym(V)}$
\end{proposition}
\begin{proof} 

All nontrivial interactions between $bv_0$ and $\Sym(V)$ parts inside 
$bv_{\mu}$ stem from $\mu(A_i)x$ part in the formula (\ref{E:pnshgrt}) 
which we kill passing from $\widehat{bv_{\mu}}$ to $\Gr\ 
\widehat{bv_{\mu}}$.

In the case at hand $H=\widehat{bv_0}$, $G=\Sym(V)$. The symmetric algebra 
$\Sym(V)$ has a diagonal $\Delta$ which on $v\in V$ is equal to 
$\Delta(v)=v\otimes1+1\otimes v$. The arguments remain to be valid in 
noncomplete and graded case.
\end{proof} 
\begin{proposition}\label{E:hrtfds}
The map $p:(\widehat{BV_{\mu}},Q)\rightarrow (\widehat{E_{\mu}YM},d)$ 
defined in equations (\ref{Jnfgfge}) is a quasiisomorphism. The map 
$p:(BV'_{\mu},Q)\rightarrow (E_{\mu}YM',d)$ is also a quasiisomorphism.
\end{proposition}
\begin{proof}

Define a filtration $\tilde{F}^nE_{\mu}YM_0$ by the formula
\begin{equation}
\tilde{F}^nE_{\mu}YM=\bigoplus_{k\geq n} \Lambda^{k}(V)\otimes.
\widehat{YM_0}
\end{equation}
 Filtrations  $\tilde{F}^n\widehat{E_{\mu}YM}$ are compatible with the map 
$p:\widehat{BV_{\mu}} \rightarrow \widehat{E_{\mu}YM}$. The map $p$ 
induces a map of spectral sequences associated with filtrations.

The term $E_0$ of the spectral sequnce associates with 
$\widehat{BV_{\mu}}$ coinsides with $\Gr \widehat{BV_{\mu}}$.

The proposition (\ref{P:Kund}) implies a series of quasiisomorphisms:
\begin{equation}\label{E:kyufg}
\widehat{\Gr BV_{\mu}}\overset{def}{=}\wBar(\Gr 
\widehat{bv_{\mu}})=\wBar(\widehat{bv_0}\otimes 
\widehat{\Sym}(V))\overset{k}{\rightarrow} \wBar(\widehat{bv_0})\otimes 
\wBar(\widehat{\Sym}(V))
\end{equation}
The map $\Gr(p)$ factors through the map $k$: $\Gr(p)=p'\circ k$. where 
$p'$ is
\begin{equation}\label{E:kykrt}
\wBar(bv_0)\otimes \wBar(\Sym(V))\overset{p'}{\rightarrow} 
\widehat{YM}_0\otimes \Lambda(V).
\end{equation}
The map $p'$ in the formula (\ref{E:kykrt}) is a tensor product of two 
quasiisomorphism. The first one is from proposition (\ref{T:fff}); the 
second one is a classical quasiisomorphism 
$\wBar(\widehat{\Sym(V)})\rightarrow \Lambda(V)$.

This considerations imply that there is an isomorphism $p:H^{\bullet}\Gr 
BV_{\mu}\rightarrow H^{\bullet} \Gr E_{\mu}YM$. It means that we have an 
isomorphism of spectral sequnces associated with filtration $\tilde{F}^n$ 
starting with $E_1$ term.

It implies that the map $p$ induces a quasiisomorphism of completed 
complexes:
$p:\widehat{BV}_{\mu}\rightarrow \widehat{ E_{\mu}YM}$.
The proof goes through in graded case. (All the tools which are needed for the
proof are collected in the Appendix in the segment devoted homogeneous
A$_{\infty}$-(co)algebras). The obstacle for the proof in 
noncomplete case is the absence of quasiisomorphism $\Barr 
bv_0=BV_0\rightarrow YM_0$.
\end{proof}

\begin{proposition}\label{P:d33q}
 $(\widehat{BV_{\mu}},Q)$, $\widehat{T_{\mu}YM}$ and $(BV'_{\mu},Q)$, 
$T_{\mu}YM'$ are pairs of quasiisomorphic algebras.
\end{proposition}
\begin{proof} 

By proposition (\ref{E:hrtfds}) the algebra $(\widehat{BV_{\mu}},Q)$ is 
quasiisomorphic to $(\widehat{E_{\mu}TYM},Q)$. By lemma (\ref{L:lknb}) the 
algebra $(\widehat{E_{\mu}TYM},Q)$ is quasiisomorphic to 
$\widehat{T_{\mu}YM}$. The proof for the second pair is similar.
\end{proof}
\begin{proposition}\label{P:mhdq}
Hochschild homology $H_i(\widehat{T_{\mu}YM},\mathbb{ C})$ as A$_{\infty}$- coalgebra 
is isomorphic to A$_{\infty}$-coalgebra $\widehat{bv_{\mu}}$. 
The same isomorphism holds in graded case.
\end{proposition}
\begin{proof} 

There is a series of quasiisomorphisms
\begin{equation}\label{E:jkrek}
\widehat{T_{\mu}YM}\overset{a}{\cong}\widehat{E_{\mu}YM}\overset{b}{\cong}\wBar
(\widehat{bv_{\mu}})
\end{equation}
A theorem (\ref{T:lskdfh}) assert that if all quasiisomorphisms in 
equation (\ref{E:jkrek}) filtered then we have a quasiisomorphism 
\begin{equation}\label{Elfgjkd}
\wBar \widehat{T_{\mu}YM} \cong \wBar \wBar (\widehat{bv_{\mu}}).
\end{equation}

Lemma (\ref{L:lknb}) asserts that quasiisomorphism $a$ is filtered. The 
proof of proposition (\ref{E:hrtfds}) shows that the quasiisomorphism $b$ 
is filtered. A proposition (\ref{L;sssfg}) claims that for any algebra 
complete with respect to canonical filtration we have a quasiisomorphism
\begin{equation}\label{Eljvksh}
\widehat{bv_{\mu}}\cong \wBar \wBar (\widehat{bv_{\mu}}).
\end{equation}

By definition homology of algebra $\widehat{T_{\mu}YM}$ is homology of the 
bar-complex $\wBar T_{\mu}YM$. By the result of \cite{Merculov} there is a 
quasiisomorphism of A$_{\infty}$-coalgebras $H(\widehat{T_{\mu}YM})$ and 
$\wBar \widehat{T_{\mu}YM}$. Quasiisomorphisms (\ref{Elfgjkd}) and 
(\ref{Eljvksh}) finish the proof.
The proof in the graded case is similar.
\end{proof}

\begin{proposition} 

The differential $Q^{\mu}_1$ in A$_{\infty}$-coalgebra 
$\widehat{bv_{\mu}}$ is defined on 
$\underline{\widehat{\Sym}(V)}$-generators  by the formulas:
\begin{equation}\label{E:djkfds}
\begin{split}
&Q^{\mu}_1(\phi_k)=0\\
&Q^{\mu}_1(A_{ i})=- \mu(A_{i})c\\
&Q^{\mu}_1(\psi^{\alpha})=0\\
&Q^{\mu}_1(c)=0\\
&Q^{\mu}_1(c^*)=\sum_{i=1}^D\mu(A_i)A^{*i}\\
&Q^{\mu}_1(A^{*m})=\sum_{i=1}^D -\mu(A_i)\mu(A_i)A_m+\mu(A_m)\mu(A_i)A_i\\
&Q^{\mu}_1(\phi^{*j})=\sum_{i=1}^D-\mu(A_i)\mu(A_i)\phi_j\\
&Q^{\mu}_1(\psi_{\alpha}^{*})=\sum_{i=1}^D\sum_{ \beta}-\Gamma_{\alpha 
\beta}^i\mu(A_i),\psi^{\beta}
\end{split}
\end{equation}%.
The same formulas hold in graded case. The homological grading on the complex $\widehat{W}$ is defined as follow:
the components of homological degree $i$ is equal to $\widehat{W}_i$.
\end{proposition}
\begin{proof} 

Direct inspection.
\end{proof}

\subsection{Examples of computations}\label{S:naiwfc}
{\bf Example 0}

The first trivial example is when $V=0$ and $\mu=0$. In this case 
differential $Q_1$ in (\ref{E:djkfds}) are equal to zero and we get that 
$\underline{H_{\bullet}(\widehat{YM_0})}=W$ where graded space $W$ is 
defined by the formula (\ref{D:kukf}). This is a tautological result.

The second example is when $dim(V)=1$. We have to options: restriction of 
the bilinear form $(.,.)$ on the kernel of the map (\ref{E:igfd}) is a. 
{\bf invertible}; b. {\bf degenerate}.

{\bf Example 1}

Let us analyse the case a.
Below is an explicit description of the complex (\ref{E:djkfds})
\begin{equation}\label{Eydhlsdd}
\mbox{
\scriptsize{
$\begin{array}{ccccccccccc}
L\otimes c^*& \overset{t}{\rightarrow}&L\otimes A^{*1} & & & &\\
 & & L\otimes A^{*2} & \overset{t^2}{\rightarrow}& L\otimes A_2& &\\
	 & & & \dots & & &\\
	 & & L\otimes A^{*D} & \overset{t^2}{\rightarrow}& L\otimes A_D& 
&\\
 & & L\phi^{*1} & \overset{t^2}{\rightarrow}& L\otimes \phi_1& &\\
 & & & \dots & & &\\
 & & L\otimes \phi^{*d'} & \overset{t^2}{\rightarrow}& L\otimes \phi_{d'}& 
&\\
	 & & L\otimes \psi^{*}_{\alpha} & \overset{t}{\rightarrow}& 
L\otimes \psi_{}^{\alpha}& &\\
	 & & & \dots & & &\\
 & & & &L\otimes A_1 &\overset{t}{\rightarrow}&L\otimes c \\\hline
 3 & & 2 & &1 & &0& &
\end{array}$
}
}
\end{equation}
where $L=\widehat{\mathbb{ C}[t]}$.
The cohomological classes of this complex are 

{\bf a} In dimension $0$ it is a space of constants.

{\bf b} In dimension $1$ the space is spanned by $A_2,\dots,A_D$, 
$tA_2,\dots,tA_D$,$\phi_1,\dots,\phi_{d'}$, $t\phi_1,\dots,t\phi_{d'}$, 
$\psi^{\alpha}$.

{\bf c} In dimension $2$ the space is spanned by $A^{*1}$.

{\bf d} In dimension $3$ the space of cocycles is zero.

This computation enables us to identify the algebra $T_{\mu}YM$ with 
subalgebra $K\subset YM_0$ defined in theorem (\ref{T:obms}). The 
connection is $[q_i]=A_{i+1}$,$[p^i]=tA_{i+1}$,$[P^j]=\phi_j$, 
$[Q^j]=t\phi_j$,$[\psi^{\alpha}]=\psi^{\alpha}$. The symbol $[a]$ denotes 
the homology class of a generator $a$. The cocycle $A^{*1}$ corresponds to 
the only relation $\sum_{i=1}^{D-1} [q_i,p^i]+\sum_{i=1}^{d'} 
[Q_i,P^i]-\frac{1}{2}\sum_{\alpha} \{\psi^{\alpha},\psi^{\alpha}\}$.

The algebra $K$ has homological dimension $2$. There is up to a constant 
only one homological class which we denote $\int \in H_2(K)$. 
The algebra $\underline{K}$ is a universal envelloping algebra of a Lie algebra
$\mathfrak{k}$ with the same set of generators and relations. We have an
 isomorphism $H^2(K)=H^2(\mathfrak{k},\mathbb{C})$.
It can be used to define a symplectic structure on moduli spaces of 
representations of $\mathfrak{k}$ in a semisimple Lie algebra $\mathfrak{ g}$, 
equipped with an invariant dot-product $(.,.)_\mathfrak{ g}$. It is well 
know what a tangent space to a point $\rho$ of the moduli of 
representations of a Lie algebra $\mathfrak{ m}$ is. It is equal to 
$H^1(\mathfrak{ m},\mathfrak{ g})$. In our case it is equal to 
$H^1(\mathfrak{k},\mathfrak{ g})$. If we have two elements $a,b\in H^1(\mathfrak{k},\mathfrak{ 
g})$, a cohomological product $a\cup b \in H^2(\mathfrak{k},\mathfrak{ g}\otimes 
\mathfrak{ g})$. A composition with $(.,.)_\mathfrak{ g}$ gives an element 
of $H^2(\mathfrak{k},\mathbb{ C})$, whose value on homological class $\int$ is equal 
to the value of the symplectic dot-product $\omega(a,b)$. In more 
condensed notations we can write:
\begin{equation}
\omega(a,b)=\int (a,b)_\mathfrak{ g}
\end{equation}
\begin{proposition} 

Symplectic form $\omega(a,b)$ defined on the moduli space $Mod_{\mathfrak{k}}(\mathfrak{ g})$ is nondegenerate and closed.
\end{proposition}
\begin{proof}

There is a different description of the space $Mod_{\mathfrak{k}}(\mathfrak{ g})$. 
Consider a linear space $(\mathfrak{ g}+ \mathfrak{ g})^{\times 
(D-1)}+(\mathfrak{ g}+ \mathfrak{ g})\otimes \Phi^* + \Pi \mathfrak{ 
g}\otimes\mathbf{S}^*$. We can identify vector space $\mathfrak{ g}+ 
\mathfrak{ g}$ with  $\mathfrak{ g}+ \mathfrak{ g}^*$, by means of 
invariant  bilinear form $(.,.)_\mathfrak{ g}$. The space $\mathfrak{ g}+ 
\mathfrak{ g}^*$ is a symplectic manifold. The space $\Pi \mathfrak{ g}$ 
is an odd-dimensional symplectic manifold with symplectic form equal to 
$(.,.)_\mathfrak{ g}$. The Lie algebra $\mathfrak{ g}$ acts on this space 
by symplectic vector fields. Define a set of functions 
$f_i=(e_i,\sum_{k=1}^{D-1} [q_k,p^k]+\sum_{k=1}^{d'} 
[Q_k,P^k]-\frac{1}{2}\sum_{\alpha} \{\psi^{\alpha},\psi^{\alpha}\})$. It 
is easy to see that this set of functions defines a set of Hamiltonians 
for generators $e_i$ of the Lie algebra $\mathfrak{ g}$. A symplectic 
reduction with respect to the action of $\mathfrak{ g}$ gives rise 
precisely the manifold we are studying. 
The statement of the proposition follows from the general properties of 
Hamiltonian reduction.
\end{proof}

{\bf Example 3}

Now we want to discuss the case b. where the restriction of the bilinear 
form on the kernel of the map $\mu$ is degenerate. It is easy to see that 
the null space of the form is one-dimensional. Without loss of generality 
we may assume that $\mu(A_0)=t$, $\mu(A_1)=it$ ($i$ is the imaginary unit) 
and the map $\mu$ on the rest of the generators is equal to zero. It is 
convenient to make a change of coordinates 
\begin{equation}\label{E:jkfnc}
\begin{split}
&v=\frac{1}{\sqrt{2}}(A_1+iA_2)\\
&u=\frac{1}{\sqrt{2}}(A_1-iA_2)\\
&u^*=\frac{1}{\sqrt{2}}(A^{1}+iA^{*2})\\
&v^*=\frac{1}{\sqrt{2}}(A^{1}-iA^{*2})
\end{split}
\end{equation}
 
In these notation the differential $Q_1$ looks like:

\begin{equation}\label{E:ydhlsdd}
\mbox{
\scriptsize{
$\begin{array}{ccccccccccc}
\mathbb{ C}[t]\otimes c^*& \overset{\sqrt{2}t}{\rightarrow}&\mathbb{ 
C}[t]\otimes u^* & & & &\\
 & & \mathbb{ C}[t]\otimes v^{*} & \overset{-2t^2}{\rightarrow}& \mathbb{ 
C}[t]\otimes v& &\\
 & & \mathbb{ C}[t]\otimes A^{*3} & \overset{0}{\rightarrow}& \mathbb{ 
C}[t]\otimes A_3& &\\
 & & & \dots & & &\\
 & & \mathbb{ C}[t]\otimes A^{*D} & \overset{0}{\rightarrow}& \mathbb{ 
C}[t]\otimes A_D& &\\
 & & \mathbb{ C}[t]\phi^{*1} & \overset{0}{\rightarrow}& \mathbb{ 
C}[t]\otimes \phi_1& &\\
 & & & \dots & & &\\
 & & \mathbb{ C}[t]\otimes \phi^{*d'} & \overset{0}{\rightarrow}& \mathbb{ 
C}[t]\otimes \phi_{d'}& &\\
 & & \mathbb{ C}[t]\otimes \psi^{*}_{\alpha} & \overset{Gt}{\rightarrow}& 
\mathbb{ C}[t]\otimes \psi^{\alpha}& &\\
 & & & \dots & & &\\
 & & & &\mathbb{ C}[t]\otimes u &\overset{\sqrt{2}t}{\rightarrow}&\mathbb{ 
C}[t]\otimes c \\\hline
 3 & & 2 & &1  & &0& &
\end{array}$
}
}
\end{equation}
where $G$ is a linear map $\mathbf{S}^*\rightarrow \mathbf{S}$. In degenerate case not much could be said about $G$. If $G$ build from spinorial $\Gamma$-matrices, $G$ has a kernel with dimension equal to $1/2dim(\mathbf{S})$. An important observation is that the complex (\ref{E:ydhlsdd}) has infinite homology groups in dimensions $1,2$. The homology in dimension $3$ is trivial and zero homology is one-dimensional. To simplify formulas for truncated Yang-Mills algebra in this case we get rid of fermions.
After change of variable (\ref{E:jkfnc}) relations (\ref{E:relssfs1}, 
\ref{E:relssfs2},\ref{E:relssfs3}) become
\begin{equation}\label{E:jkfvdnm}
\begin{split}
&-[v[v,u]]+\sum_{i=3}^D[A_{i},[A_{i},v]]+\sum_{k=1}^{d'}[\phi_k[\phi_k,v]]=0 
\notag \\
&-[u[u,v]]+\sum_{i=3}^D[A_{i},[A_{i},u]]+\sum_{k=1}^{d'}[\phi_k[\phi_k,u]]=0
\notag \\
&[u[v,A_m]]+[v[u,A_m]]+\sum_{i=3}^D[A_{i},[A_{i},A_{m}]]+\sum_{k=1}^{d'}[\phi_k[\phi_k,A_{m}]]\notag \\
&=0 \quad m=3,\dots,D \\
&[u[v,\phi_j]]+[v[u,\phi_j]]+\sum_{k=3}^D[A_k[A_k.\phi_j]]+\frac{\partial 
U}{\partial \phi_j}=0 \quad j=1,\dots,d'
\end{split}
\end{equation}
The generators of the algebra $T_{\mu}YM$ with $rk(\mu)=1$ and 
$ind(\mu)=0$ are 
\begin{equation}
\begin{split}
&v\\
&p=[u,v]\\
&A^n_m=Ad^n(u)A_m \quad n\geq 0, m=3,\dots D\\
&\phi_j^n=Ad^n(u)\phi_j \quad n\geq 0, j=1,\dots d'
\end{split}
\end{equation}. As in the case of the first example the algebra $YM_0$ is 
a semidirect product of an abelian one-dimensional Lie algebra and algebra 
$T_{\mu}YM$. The relations in $T_{\mu}YM$ and the action of the generator 
of the abelian Lie algebra (Hamiltonian) can be read off from equations 
(\ref{E:jkfvdnm}). The action of the Hamiltonian is given by the formulas:
\begin{equation}
\begin{split}
&H(p)=\sum_{i=3}^D[A^0_{i},A^1_{i}]+\sum_{k=1}^{d'}[\phi^0_k,\phi^1_k,] 
\notag \\
&H^m(A_{i}^0)=A_{i}^{m}\\
&H^m(\phi_{i}^0)=\phi_{i}^{m}\\
\end{split}
\end{equation}
The relations are:
\begin{equation}\label{R:llaqw}
\begin{split}
&[v,p]+\sum_{i=3}^D[A^0_{i},[A^0_{i},v]]+\sum_{k=1}^{d'}[\phi^0_k[\phi^0_k,v]]=0 
\notag \\
&A^{*m}_0=[p,A^0_m]+2[v,A^1_m]+\sum_{i=3}^D[A^0_{i},[A^0_{i},A^0_{m}]]+\sum_{k=1}^{d'}[\phi^0_k[\phi^0_k,A^0_{m}]]=0 
\notag \\
&\quad m=3,\dots,D \\
&\phi^{*j}_0=[p,\phi^0_j]+2[v,\phi^1_j]+\sum_{k=3}^D[A^0_k[A^0_k.\phi^0_j]]+\frac{\partial 
U}{\partial \phi^0_j}=0 \quad j=1,\dots,d'\\
&A_n^{*m}=H^n(A_0^{*m})\quad m=3,\dots,D\quad n\geq 1\\
&\phi_n^{*j}=H^n(\phi_0^{*j})\quad j=1,\dots,d'\quad n\geq 1
\end{split}
\end{equation}

\begin{proposition}\label{P:yghlf}
There is an isomorphism of $T_{\mu}YM$ and a quotient algebra $\mathbb{ 
C}<v,p,A^n_k,\phi^n_j,\psi_{\alpha}>/(I)$ where ideal is generated by 
relations (\ref{R:llaqw}). There is an isomorphism $\mathbb{ C}[H]\ltimes 
T_{\mu}YM \cong YM_0$.
\end{proposition}

{\bf Example 4.}

Suppose $V=\mathbf{ V}$ and $\mu=id$. 
We need to compute cohomology of complexes:

\begin{equation}\label{E:ydhld}
\mbox{
\scriptsize{
$\begin{array}{c}\end{array}\begin{array}{ccccccccccc}
\begin{array}{c} \\ <c^*>\otimes \underline{\widehat{\Sym}(\mathbf{ 
V})} \\ \\ \end{array} &
\begin{array}{c} \\ \underset{\rightarrow}{d} \\ \\ \end{array} &
\begin{array}{c}\mathbf{ S}^*\otimes \underline{\widehat{\Sym}(\mathbf{ 
V})}\\ \\ \Pi \mathbf{ V}^*\otimes \underline{\widehat{\Sym}(\mathbf{ V})} 
\\ \\ \Phi^{*}\otimes \underline{\widehat{\Sym}(\mathbf{ V})} \end{array}&
\begin{array}{c}\underset{\rightarrow}{ \mbox{$\dirac$}}\\ 
\\\underset{\rightarrow}{d*d} \\ \\ \underset{\rightarrow}{\times ||v||^2} 
\end{array}&
\begin{array}{c}\mathbf{ S}\otimes \underline{\widehat{\Sym}(\mathbf{ 
V})}\\ \\ \mathbf{ V}\otimes \underline{\widehat{\Sym}(\mathbf{ V})} \\ \\ 
\Phi \otimes \underline{\widehat{\Sym}(\mathbf{ V})}\end{array}&
\begin{array}{c} \\ \underset{\rightarrow}{d} \\ \\ \end{array}&
\begin{array}{c} \\ <c>\otimes \widehat{\Sym}(\mathbf{ V}) \\ \\ 
\end{array} &
\\\hline
3 & & 2 & &1 & &0
\end{array}$
}
}
\end{equation}
It is easy to see that homology of the complex (\ref{E:ydhld}) coincide with 
completion of homology of a similar complex with $\underline{\widehat{\Sym}}$
stripped off the completion sign. Therefore we will examine only noncompleted version.
It is particularly easy to compute $\Phi^{*}-\Phi$ part of cohomology. It 
is equal zero in all dimensions but one where it is $\Phi \otimes 
\Sym(\mathbf{ V})/(||v||^2)$. By $(||v||^2)$ we denote a homogeneous ideal 
of functions equal to zero on quadric $q$ given by equation $||v||^2=0$.

Denote
\begin{equation}
\begin{split}
& T = \{ (a,b)\in \mathbf{ V}\times \mathbf{ V}|||a||^2=0,(a,b)=0\}\\
& X = \{ (a,\tilde{b})\in T| b \mbox{ is defined up to addition of 
multiple of } a\}.
\end{split}
\end{equation}
Then $X$ is a quotient bundle of the tangent bundle $T$ to the quadric by 
one-dimensional subbundle $L$. $L$ consists of all vector fields that are 
tangent to the projection $q\{0\}\rightarrow \tilde{q}\subset \mathbb{ 
P}^{D-1}$. The space of global sections of $X$ is precisely the first 
cohomology of the complex (\ref{E:ydhld}) in $\mathbf{ V}\otimes 
\Sym(\mathbf{ V})$-term. The zero cohomology in $<c>\otimes \Sym(\mathbf{ 
V})$ term is one-dimensional by obvious reason. Vanishing of the third and 
the second cohomology will be proved in proposition (\ref{P:kfiuw}) under 
more general assumptions.

There is a standard "adjoint" Dirac operator $\dirac^*:\mathbf{ S}\otimes 
\Sym(\mathbf{ V})\rightarrow \mathbf{ S}^{*}\otimes \Sym(\mathbf{ V})$. 
Together $\dirac$ and $\dirac^*$ satisfy 
\begin{equation}\label{H:kand}
\begin{split}
&\dirac\dirac^*=|| . ||^2\\
&\dirac^*\dirac=|| . ||^2
\end{split}
\end{equation} where $|| . ||^2$ is an operator of multiplication on 
quadric. Equations (\ref{H:kand}) imply that there is no second cohomology 
in $\mathbf{ S}^*\otimes \Sym(\mathbf{ V})$-term. There is a similar 
geometric interpretation of cohomology in $\mathbf{ S}\otimes 
\Sym(\mathbf{ V})$-term. Suppose $\mathbf{ S}$ is a spinor representation 
of $Spin(n)$, upon restriction of $\mathbf{ S}$ onto $Spin(n-2)$ $\mathbf{ 
S}$ splits into two nonisomorphic spinor representation $\mathbf{ 
S}_1,\mathbf{ S}_2$, choses the one from the two which contains the 
highest vector of $\mathbf{ S}$ as representation of $Spin(n)$. The Levi 
subgroup of stabilizer of a point $l$ of the quadric $q$ is equal to 
$SO(n-2)$. One can induce a vector bundle $C$ on $q$ from representation 
$\mathbf{ S}_1$ of $Spin(n-2)$. It is not hard to see that first 
cohomology of complex (\ref{E:ydhld}) in term $\mathbf{ S}\otimes 
\Sym(\mathbf{ V})$ is isomorphic to direct sum of the space of global 
section of $C$.

It is useful to use Borel-Weyl theorem to compute the spaces of global 
section of these bundles. 

As an illustration let us carry out such computation in the case 
$D=10$,$d'=0$,$N=1$.

The Dynkin graph of the group $Spin(10)=Spin(\mathbf{ V})$ is
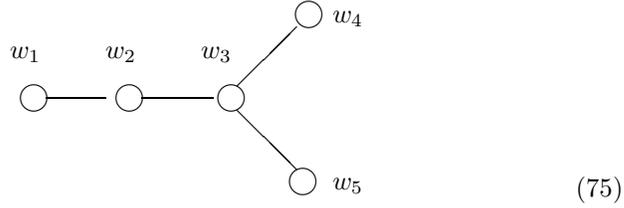
\begin{equation}\label{P:kxccvg}
\mbox{
\setlength{\unitlength}{3947sp}
\begin{picture}(2025,1246)(301,-452)
\thinlines
\put(1688,164){\circle{168}}
\put(451,164){\circle{168}}
\put(1051,164){\circle{168}}
\put(2176,689){\circle{168}}
\put(2138,-361){\circle{168}}
\put(1126,164){\line( 1, 0){450}}
\put(1726,239){\line( 1, 1){375}}
\put(1726, 89){\line( 1,-1){375}}
\put(526,164){\line( 1, 0){375}}
\put(301,389){\makebox(0,0)[lb]{$w_1$}}
\put(901,389){\makebox(0,0)[lb]{$w_2$}}
\put(1501,389){\makebox(0,0)[lb]{$w_3$}}
\put(2326,-436){\makebox(0,0)[lb]{$w_5$}}
\put(2326,614){\makebox(0,0)[lb]{$w_4$}}
\end{picture}
}
\end{equation}
We will encode a representation which is labeled by Dynkin diagram above 
by an array
 $[w_1,w_2,w_3,w_4,w_5]$.
Our convention is that spinor representation $\mathbf{ S}$ is equal to 
irreducible representation with highest weight $[0,0,0,1,0]$, the 
tautological representation in $\mathbb{ C}^{10}=\mathbf{ V}$ is equal to 
$[1,0,0,0,0]$, the exterior square of the later representation is equal to 
$[0,1,0,0,0]$. 
 In case of $N=1,D=10$ super Yang-Mills theory the cohomology are equal to 
\begin{equation}
\begin{split}
&\bigoplus_{i\geq 0}[i,1,0,0,0] \quad \mbox{ harmonic two-forms}\\
&\bigoplus_{i\geq 0}[i,0,0,1,0] \quad \mbox{ harmonic spinors}
\end{split}
\end{equation}

An interesting feature of the algebra $T_{\Id}YM$ is that its second 
homology vanish. As a result we conclude that the algebra 
$\widehat{T_{\Id}YM}$ is a completed free algebra. It is useful to exhibit 
the set of free generators of such algebra. Before doing this introduce 
some notations. The space $\mathbf{ V}$ is a representation of $SO(10)$ 
and a basis vector $A_D$ can be considered as the highest vector. The 
element $(A_D)^{\otimes i}$ is a highest vector in the i-th symmetric power 
$\Sym^i(\mathbf{ V})$ . Let $W$ be an irreducible representation of 
$Spin(D)$ with highest vector $w$. Then a vector $(A_D)^i\otimes w \in 
\Sym^i(\mathbf{ V})\otimes W$ will be a highest vector and generates an 
irreducible subrepresentation of $Spin(D)$ in $\Sym^i(\mathbf{ V})\otimes 
W$ denote the projection on such representation by $p$. For example 
representation $[i,1,0,0,0]$ is isomorphic to the image of 
$p:\Sym^i(\mathbb{ C}^{10})\otimes \Lambda^2(\mathbb{ C}^{10})\rightarrow 
\Sym^i(\mathbb{ C}^{10})\otimes \Lambda^2(\mathbb{ C}^{10})$. 

Denote $Ad(x)(y)=[x,y]$, $F_{ij}=[A_i,A_j]$. Introduce an elements 
$Ad(A_{(i_1})\dots Ad(A_{i_{k-1}})F_{i_{k})j}$ where $()$ denotes 
symmetrization. This element belongs to $\Sym^{k-1}(\mathbf{ V})\otimes 
\Lambda^2(\mathbf{ V})$. Similarly elements $Ad(A_{(i_1})\dots 
Ad(A_{i_{k}})\phi_j$ belong to $\Sym^{k}(\mathbf{ V}) \otimes \Phi$ and 
$Ad(A_{(i_1})\dots Ad(A_{i_{k-1}})\psi^{\alpha}$ belong to 
$\Sym^{k}(\mathbf{ V})\otimes \mathbf{ S}$. The elements 
\begin{equation}
\begin{split}
&p(Ad(A_{(i_1})\dots Ad(A_{i_{k-1}})F_{i_{k})j})\\
&p(Ad(A_{(i_1})\dots Ad(A_{i_{k})})\psi^{\alpha})\\
&p(Ad(A_{(i_1})\dots Ad(A_{i_{k})})\phi_j)
\end{split}
\end{equation}
is a topological free set of generators of the algebra $T_{\Id}YM$. One 
can check it by looking at the image of these elements in the first 
homology. 

We would like to elucidate some general features of the complex 
(\ref{E:djkfds}). Its structure depends on the map $\mu$. 
\begin{definition}
Let $\mu(b)$ be the image of the tensor $q\in \Sym^2(\mathbf{V})$ inverse to the scalar product under the map $\mu$ (see \ref{E:igfd}). Denote $ind(\mu)=rk\mu(b)$ and $rk(\mu)=dim V$.
\end{definition}
There are three classes of maps:\ 
\\
{\bf a.} $\mu=0$

{\bf b.} $ind(\mu)=0,\mu \neq 0$

{\bf c.} $ind(\mu)=1$, $rk(\mu)=1$.

{\bf d.} All other cases

The importance of such division is justified by 
the following Proposition:
\begin{proposition}\label{P:kfiuw}\ 

{\bf a.} If condition a. is satisfied the algebra $T_{\mu}YM$ has homological 
dimension $3$ and coinsides with $YM_0$.

{\bf b.} If condition b. is satisfied the algebra $T_{\mu}YM$ has homological 
dimension $2$ has infinite number of generators and relations.

{\bf c.} If condition c. is satisfied then the algebra $T_{\mu}YM$ has 
homological dimension $2$ , has finite number of generators and one 
relation.

{\bf d.} If the condition d. is satisfied the algebra $T_{\mu}YM$ is a completed 
free algebra, with infinite number of generators.
\end{proposition}
\begin{proof}\ 

{\bf a.} Proof is tautology. 

If $\mu\neq 0$ the the third homology is equal to zero. Indeed the space 
of four-chains is equal to zero. On the space of three-chains (equal to 
$\Sym(V)\otimes c^*$) the differential is injective. It implies that 
if $\mu\neq 0$ the homological dimension of all algebras in question is 
less or equal to two.

{\bf b.} The case when dimension $d$ (the number of generators $A_k$) is 
less or equal to three and $ind(\mu)=0$ was covered by example 2. We may 
assume that $d\geq 4$. Then the restriction of the map $Q_1: \Pi \mathbf{ 
V}^*\otimes \widehat{\Sym}(V)\rightarrow \mathbf{ V}\otimes \widehat{\Sym}(V)$
 contains a free 
module with at least $[d/2]$ generators ($[d/2]$ is the dimension of 
maximal isotropic subspace in a space $\Pi \mathbf{ V}^*$ equipped with 
the standard bilinear form). It implies that the image 
$Im[Q_1:\overline{W}_3\rightarrow \overline{W}_2]$ could not cover $Ker 
Q_1\cap \overline{W}_2$ and the second cohomology are 
infinite-dimensional.

{\bf c.} This was worked out as a first nontrivial example of computation 
of cohomology.

{\bf d.} If $ind(\mu)\geq 1$ one can choose a subspace $V'\subset V$ of 
codimension $1$ such that the image of $\mu(b)$ in $\Sym^2(V/V')$ is 
nonzero. Choose some complement $V''$ to $V'$ such that $V=V'+V''$.
 Define a linear space $F^1$ as an ideal in $\widehat{\Sym}(V)$ generated by 
$V''$.  Introduce a multiplicative descending filtration $F^i$ of 
$\widehat{\Sym}(V)$ generated by $F^1$. Define a filtration of $\widehat{\overline{W}}$
as $F^s(\widehat{\overline{W}})=W\otimes F^s$. The $E_2$ term of a spectral sequence 
associated with such filtration is equal to cohomology of 
(\ref{Eydhlsdd}) where $L=\widehat{\mathbb{ C}[t]}\hat{\otimes} \widehat{\Sym}(V')$.
 The element $t$ 
is a generator of $V''$. The space of second homology of $E_2$-term is a 
free $\widehat{\Sym}(V')$-module generated by $A^{*1}$ (we can assume that without 
loss of generality). Similarly the first homology of $E_2$ term is a free 
$\widehat{\Sym}(V')$ module of rank $D+d'-1+l$. The differential in $E_2$ is a 
$\widehat{\Sym}(V')$ homomorphism and it is injective on second homology, if we can 
prove that it is nonzero on $A^{*1}$. A simple analysis shows that 
$Q_1A^{*1}\neq0 $ in $E_2$ if the map $\mu$ satisfies condition $d$.
\end{proof}

\subsection{Fano manifolds}\label{F:FFF}
Let $M$ be a smooth  manifold of dimension $n$. Denote 
$\Omega^p$ a sheaf of holomorphic $p$-forms on $M$. Let us fix a line
bundle  ${\cal L}$. In most interesting situations $M$ is a projective
manifold and ${\cal L}$ is obtained from tautological line bundle ${\cal 
O}(1)$ on projective space by means of restriction to $M$. We will use 
 the notation  
${\cal O}(1)$ for ${\cal L}$,  and  ${\cal O}(-1)$ 
for the dual bundle ${\cal L}^*$ also in general case. We identify line
bundles with invertible sheaves.. 
For any sheaf ${\cal F}$ 
denote ${\cal F}(i)$ the sheaf ${\cal F}\otimes {\cal O}(1)^{\otimes i}$ 
where the tensor product is taken over the structure sheaf ${\cal O}$. 

 Serre algebra ${\cal S}$ is defined by the formula
\begin{equation}\label{D:werty}
{\cal S}=\bigoplus_{i=0}^{\infty}{\cal 
S}_i=\bigoplus_{i=0}^{\infty}H^0(M,{\cal O}(i)).
\end{equation}

It can be embedded in the differential algebra ${\cal B}$ (Koszul- Serre 
algebra) in the following way.
As an algebra 
\begin{equation}\label{L:dfghs}
{\cal B}={\cal S}\otimes \Lambda({\cal S}_1),
\end{equation} where 
$\Lambda({\cal S}_1)$ is the exterior algebra of ${\cal S}_1$. The algebra 
$\cal B$ is $\mathbb{ Z}$ graded: an element $a \in {\cal S}_i\otimes 
\Lambda^j({\cal S}_1)$ has degree $deg(a)=j$. Let $v_{\alpha},\  \alpha 
=1,\dots, s$ be a basis of ${\cal S}_1\subset {\cal S}$ and 
$\theta_{\alpha}$ be the corresponding basis of ${\cal S}_1\subset 
\Lambda({\cal S}_1)$. 
The algebra $\cal B$ carries a differential $d$ of degree $-1$. If $a \in 
{\cal S}$ then $d(a)=0$, $d(\theta_{\alpha})=v_{\alpha}$. It can be 
extended to $\cal B$ by the Leibniz rule.

There is an additional grading on $\cal B$ which we denote by $Deg$. An 
element $a \in {\cal S}_i\otimes \Lambda^j({\cal S}_1)$ has degree 
$Deg(a)=i+j$. The differential has degree zero with respect to additional 
grading. According to definition in Appendix such algebra is called homogeneous.
We can split $ \cal B $ into a sum 
\begin{equation}
{\cal B}=\bigoplus_{i,j}{\cal B}_{j,i}
\end{equation}
such that $a\in {\cal B}_{j,i}$ has the degrees $deg(a)=j,Deg(a)=i$. 
A line bundle ${\cal L}$ is called ample if for some positive $n$ the tensor power ${\cal L}^{\otimes n}$ defines an embedding of the manifold $M$ into $\mathbf{P}(H^0(M,{\cal L}^{\otimes n})^*)$. 
We assume that
\begin{equation}\label{A:gfd}
\begin{split}
&M \mbox{ is an algebraic smooth  manifold of dimension } n, \\
&\mbox{ canonical bundle } \Omega^{n} \mbox{ is isomorphic to } {\cal 
O}(-k)\quad k>0, \\
&\mbox{\cal O}(1)={\cal L} \mbox{ is ample .}\\
\end{split}
\end{equation}
The constant $k$ is called index.
 Manifolds satisfying
 above assumptions are  Fano manifolds (i.e. the anticanonical line 
bundle is ample). Conversely, every Fano manifold can be equipped with
the above structure. (We can always take $k=1$.)
 
 The goal of this section  is to illuminate some properties of cohomology 
of the following complexes :

\begin{equation}
Q^{\bullet}_i=\left( 0 \rightarrow \Lambda^{i} ({\cal S}_1) \rightarrow 
\Lambda^{i-1} ({\cal S}_1) \otimes {\cal S}_{1}\rightarrow \dots 
\rightarrow {\cal S}_{i}\rightarrow 0\right)=\bigoplus_{j=0}^i{\cal B}_{j,i}
\end{equation}

{\bf Some preliminaries on $H^{\bullet}(M,{\cal O}(i))$} 
\begin{proposition}\label{T:dsa}(Kodaira)\cite{Griffits}
Suppose ${\cal L}$ is an ample bundle over any complex  manifold $N$. Then 

$H^i(N,\Omega^j\otimes {\cal L})=0$ if $i+j>n$.

\end{proposition}
\begin{corollary}\label{C:kighg} 
In assumptions \ref{A:gfd}
\begin{equation}\label{E:lij}
\begin{split}
&H^{i}(M,{\cal O}(l))=0 \mbox{ for } 0<i<n \mbox{ and any } l\\
&H^{0}(M,{\cal O}(-l))=0 \mbox{ for } l>0\\
&H^{0}(M,{\cal O}(l))=0 \mbox{ for } l>-k\\
&H^{0}(M,{\cal O})=\mathbb{ C} \\
&H^{n}(M,{\cal O}(-l))=H^{0}(M,{\cal O}(l-k))^*\quad l\geq k
\end{split}
\end{equation}

\end{corollary}
\begin{proof} 

The proof is straightforward: use theorem (\ref{T:dsa}) and Serre duality.
\end{proof}

\begin{theorem}\label{T:dli}
Suppose $M$ is an  $n$ dimensional Fano manifold of index $k$. Let ${\cal B}$ be the differential algebra (\ref{L:dfghs}). 
There exists  a nondegenerate pairing
$$H^{j,i}({\cal B})\otimes H^{s-n-1-j,s-k-i}({\cal B})\rightarrow 
H^{s-n-1,s-k}({\cal B})$$
where $s={\rm dim}{\cal S}_1$.
\end{theorem}
\begin{proof}

%Consider a projective space $\mathbb{ P}({\cal S}^*_1)$. 
There is a short 
exact sequence of vector bundles over it(Euler sequence):
\begin{equation}\label{E:jdaolqs}
0 \rightarrow {\cal O}(-1)\rightarrow {\cal S}^*_1 
\overset{m^*}{\rightarrow} T(-1)\rightarrow 0
\end{equation}

%Where $T_{\mathbb{ P}({\cal S}^*_1)}$ is the tangent bundle on $\mathbb{ 
%P}({\cal S}^*_1)$,
Where  ${\cal S}^*_1$ is a trivial bundle with a fiber ${\cal S}^*_1$. The first map is tautological embedding. the vector bundle $T(-1)$ is a quotient ${\cal S}^*_1/{\cal O}(-1)$
%We are going to use the same notations for the pullback (\ref{E:jdaolqs}) on $M$ via the map 
%$\psi:M\rightarrow \mathbb{ P}({\cal S}^*_1)$. 

The $i$-th exterior power 
of the dual to that sequence gives rise to the complex of vector bundles

\begin{equation}
N_i=\left(0\rightarrow E^i\rightarrow \Lambda^i ({\cal S}_1) \rightarrow 
\Lambda^{i-1} ({\cal S}_1)(1) \rightarrow ... \rightarrow \Lambda^{1} 
({\cal S}_1)(i-1) \rightarrow {\cal O}(i)\rightarrow 0\right)
\end{equation}

which is acyclic everywhere except in zero degree. The zeroth cohomology 
is equal to $\Lambda^i( T(-1))=E^i$.

Since $E^s=0$ the complex $N_s$ has a particularly simple form.

After tensoring the resolution $N^{\bullet}_s(i)$ on Dolbeault complex 
$\Omega^{0,\bullet}$ one can compute diagonal cohomology of corresponding 
bicomplex ( hypercohomology ) which we denote by $\mathbb{ 
H}^{\bullet}(N_s^{\bullet}(i))$. By acyclicity of $N^{\bullet}_s(i)$ we 
have an equality
\begin{equation}\label{E:fdsfa}
\mathbb{ H}^{\bullet}(N_s^{\bullet}(i))=0
\end{equation}

 There is a spectral sequence of the bicomplex 
$N_s^{\bullet}(i)\otimes\Omega^{0,\bullet}$ whose $E_1^{p,q}$ term 
coincides with 
\begin{equation}\notag
E_1^{p,q}=\Lambda^{p} ({\cal S}_1)\otimes H^q({\cal O}(i-p)).
\end{equation}

 According to equation (\ref{E:lij}) and corollary (\ref{C:kighg}) all 
nonzero entries of $E_1$ term are concentrated on horizontal segments:
\begin{align}
&E_1^{p,0}=\Lambda^{p} ({\cal S}_1)\otimes {\cal S}_{i-p}, \ 0 \leq p 
\leq  
i \label{E:pdfgt} \\
&E_1^{p,n}=\Lambda^{p} ({\cal S}_1)\otimes {\cal S}^*_{p-i-k}, \ i+k\leq 
p 
\leq s  
\end{align}
All entries $E_1^{p,q}$ not mentioned in the above table are equal to 
zero.

\begin{definition}
Let $K^{\bullet}=\dots \rightarrow K^i\rightarrow K^{i+1}\rightarrow 
\dots$. The complex $K^{\bullet}[l]$ is a complex with shifted grading:
$$K^i[l]=K^{i+l}.\mbox{ The differential in the new complex is equal to } (-1)^{l}d.$$
\end{definition}
We have an obvious equality of complexes:
\begin{align}
&(E_1^{\bullet,0},d)=Q^{\bullet}_{i}\\
&(E_1^{\bullet,n},d)=Q^{\bullet *}_{s-k-i}[-s]
\end{align}
The second isomorphism depends on a choice of a linear functional on the 
space $\Lambda^{s}({\cal S}_1) \cong \mathbb{ C}$.
Symbol $*$ means dualisation. 
The spectral sequence converges to zero in the $n+1$-th term. The 
differential $d_{n+1}$ on $E^{p,q}_2=E^{p,q}_{n+1}$ induces an isomorphism 
\begin{equation}\label{E:cvjkdhn}
d_{n+1}:E^{p,n}_2 \rightarrow E^{p-n-1,0}_2
\end{equation}
which is a map of $E^{\bullet,0}_2$ modules, because the spectral sequence 
is multiplicative.

The isomorphism \ref{E:cvjkdhn} can be interpreted as  nondegenerate pairing:
\begin{equation}\label{E:ddd}
(.,.):H^l(Q_{i})\otimes H^{s-n-1-l}(Q_{s-k-i})\rightarrow \mathbb{C}
\end{equation}
It satisfies $(ab,c)=(a,bc)$, because $d_{n+1}$ is a map of $E^{\bullet,0}_2$ modules. The pairing  can be recovered from the functional 
$\lambda(a)=(a,1)$ by the rule $(a,b)=\lambda(ab)$. The functional is not 
equal to zero only on $H^{s-n-1}(Q_{s-k})$. The proof follows from this.

A direct inspection of a complex $Q_i^{\bullet}$ shows that 
$H^i(Q_i^{\bullet})=0$ for $i\neq 0$ . It implies by duality (\ref{E:ddd}) 
that 

\begin{equation}
H^{i,i}({\cal B})=H^{s-n-1-i,s-k-i}({\cal B})=0 \mbox{ for }i\neq 0.
\end{equation}
\end{proof}

Duality (\ref{E:ddd}) also implies that cohomology of $Q_{i}^{\bullet}$ 
are not equal to zero only in the range $0 \leq deg \leq s-n-1$, therefore 
we prove the following Proposition:
\begin{proposition}\label{T:asfbh}
In the assumptions of Theorem (\ref{T:dli})
\begin{align}
&H^{j,i}({\cal B})\neq 0 \mbox{ only for }\notag \\
& 0 \leq j \leq s-n-1\mbox{ and } j \leq i \notag \\
&\mbox{ and by duality } i \leq j+ n+1-k. \notag 
\end{align}%.
\end{proposition}
%\begin{remark}
%Suppose $-\Omega^n={\cal L}^{\otimes k}$ $k>0$. In order for construction of Theorem
% (\ref{T:dli}) to hold we don't need a very ampleness of ${\cal L}$ (i.e global sections 
% of ${\cal L}$ define an embedding) but only ampleness (i.e global sections of some power of ${\cal L}$ 
%define an an embedding)
%Indeed in this case the key prerequisite  (\ref {C:kighg}) still holds and proposition (\ref{T:dli}) is still valid.
%\end{remark}

Notice that an analog of Theorem (\ref{T:dli})
 can be proved for
 Calabi-Yau 
manifolds.

\begin{proposition} 
Suppose an $n$ dimensional smooth algebraic manifold has $\Omega^n={\cal O}$,
%Suppose index $k=0$ and $M,\mathcal{O}(1)$ satisfy (\ref{A:gfd})and $M$ is Calabi-Yau manifold,
 $H^i(M,\mathcal{O})=0$ for $0<i<n$ and ${\cal L}={\cal O}(1)$ is ample. Then there is a nondegenerate pairing
$$H^{j,i}({\cal B})\otimes H^{s-n-1-j,s-i}({\cal B})\rightarrow H^{s-n-1,s}({\cal B}).$$%.
for differential graded algebra ${\cal B}$ defined in (\ref{L:dfghs})
\end{proposition}
\begin{proof} 

The proof goes along the same lines as of (\ref{T:dli})
\end{proof}
\subsection{Berkovits algebra}\label{S:nhkshw}

In this section we will be dealing with some structures made of 
$16$-dimensional spinor representation $\mathbf{ S}=\mathfrak{ s}_l$ of 
group $Spin(10)$ defined over complex numbers. This group is a double 
cover of a group of all linear transformations of linear space $\mathbf{ 
V}$, $dim(\mathbf{ V})=10$ that preserve a nondegenerate form $(.,.)$ and 
have determinant equal to one. The Dynkin diagram $D_5$ that corresponds 
to Lie algebra $SO(\mathbf{ V})$ could be found on the picture 
(\ref{P:kxccvg}). Our convention is that representation $\mathbf{ S}$ is 
equal to irreducible representation with highest weight $[0,0,0,1,0]$.

Let ${\cal S}_{i}$ be $[0,0,0,i,0]$. There is a structure of algebra on 

\begin{equation}\label{E:asvdfd}
\bigoplus_{i\geq 0} {\cal S}_{i}={\cal S}
\end{equation} induced by tensor product of 
representation and projection on the leading component.

According to Cartan \cite{EC} there is a compact K\"{a}hler 
$10$-dimensional homogeneous space $OGr(5,10)$ of the group $O(\mathbf{ 
V})$-the Grassmanian of maximal ( five-dimensional) isotropic subspaces of 
$\mathbf{ V}$. This space is called isotropic Grassmanian. It has two 
connected components. They are isomorphic as K\"{a}hler manifolds but not 
as homogeneous spaces. An element $e \in O(\mathbf{ V})$ with $det(e)=-1$ 
swaps the spaces. Let us describe one of the connected components which we 
denote ${\cal Q}$ and will call a space of pure spinors. Fix $W_0 \in 
OGr(5,10) $, then another isotropic subspace $W_1$ belongs to the same 
component ${\cal Q}$ if $dim(W_0\cap W_1) $ is odd.

The complex group $Spin(\mathbf{ V})$ (in fact $SO(\mathbf{ V})$) acts 
transitively on ${\cal Q}$ ; corresponding stable subgroup $P$ is a 
parabolic subgroup. To describe the Lie algebra $\mathfrak{p}$ of $P $ we 
notice that the Lie algebra $\mathfrak{so}(\mathbf{ V}) $ of $SO(\mathbf{ 
V}) $ can be identified
with ${\Lambda}^2(\mathbf{ V}) $ (with the space of antisymmetric tensors 
${\rho}_{ab} $ where $a, b=0, \dots, 9 $). The vector representation 
$\mathbf{ V}$ of $SO(\mathbf{ V}) $ restricted to the group $GL(5, 
\mathbb{C})=GL(W) \subset SO(\mathbf{ V}) $ is equivalent to the direct 
sum $W \oplus W^* $ of vector and covector representations of $GL(W) $, 
where $dim(W)=5$. The direct sum $W+W^*$ carries a canonical symmetric 
bilinear form. The Lie algebra of $SO(\mathbf{ V}) $ as vector space can 
be decomposed as ${\Lambda}^2(W)+ \mathfrak{p} $ where $\mathfrak{p }=(W 
\otimes W^*) \oplus \Lambda^2(W^*) $
is the Lie subalgebra of $\mathfrak{p}$. Using the language of generators 
we
can say that the Lie algebra $\mathfrak{so}(10, \mathbb{C}) $ is generated 
by
skew-symmetric tensors $m_{ab},\quad n^{ab} $ and by $k_a^b $ where $
a, b=1, \dots, 5 $. The subalgebra $\mathfrak{p}$ is generated by $k_a^b $
and $n^{ab} $. Corresponding commutation relations are 
\begin{align}
& [m, m^{\prime}]=[n, n^{\prime}]=0 \\
&[m, n]_a^b=m_{ac}n^{cb} \\
&[m, k]_{ab}=m_{ac}k_b^c+m_{cb}k_a^c \\
&[n, k]_{ab}=n^{ac}k_c^b+n^{cb}k_c^a
\end{align}

\begin{proposition}(Borel-Weyl-Bott theorem \cite{Bott}) Suppose $\mathcal{L}$ is an 
invertible homogeneous line bundle over K\"{a}hler compact homogeneous space $M$ of a semisimple group $G$. Then 
$H^i(M,\mathcal{L})$ could be nonzero only for one value of $i$. 
For this value $H^i(M,\mathcal{L})$ is an irreducible 
representation. 
\end{proposition}
\begin{corollary}
$H^i(\mathcal{Q},\mathcal{O})=0$ for $i>0$.
\end{corollary}
Since $H^1(\mathcal{Q},\mathcal{O})=0$ all holomorphic topologically trivial line 
bundles are holomorphically trivial.
The corollary and Hodge decomposition imply that 
$H^2(\mathcal{Q},\mathcal{O})=H^0(\mathcal{Q},\Omega^2)=0$ and 
$H^2(\mathcal{Q},\mathbb{Z})=Pic(\mathcal{Q})$. 
\begin{proposition}\ 

{\bf a.} The group $Pic({\cal Q})=\mathbb{ Z}$.

{\bf b.} The group $Pic({\cal Q})$ has an very ample generator which we denote by 
${\cal O}(1)={\cal L}$ such that $H^0(\mathcal{Q},\mathcal{O}(1))=\mathbf{ S}$.

{\bf c.} The canonical class of $\mathcal{Q}$ is isomorphic to $\mathcal{O}(-8)$.

\end{proposition}
\begin{proof} 

We saw that that the Levi subgroup of the parabolic group $P$ contains a 
center isomorphic to $\mathbb{ C}^{\times}$, so singular cohomology 
$H^1(P,\mathbb{ Z})=\mathbb{ Z}$. Transgression arguments imply that 
$H^2({\cal Q},\mathbb{ Z})=\mathbb{ Z}$. It proves that the Picard group 
of $Q$ is equal to $\mathbb{ Z}$ .

Denote $\widetilde{GL(W)}\subset Spin(\mathbf{ V})$ a double cover of 
$GL(W)$. A restriction of $\mathbf{ S}^*$ on $\widetilde{GL(W)}$ is 
isomorphic to 
\begin{equation}
\left[\mathbb{ C}+\Lambda^2(W)+\Lambda^4(W) \right] \otimes det^{-1/2}(W).
\end{equation}
By Borel-Weyl theorem it implies that the ample generator of $Pic({\cal 
Q})$ which we denote by ${\cal O}(1)$ has a space of global sections 
isomorphic to $\mathbf{ S}$.

Consider a representation of $\widetilde{GL(W)}$ in $\Lambda^2(W)$. It is 
easy to see that 
\begin{equation}
det(\Lambda^2(W))=det^4(W)=(det^{1/2}(W))^8.
\end{equation}
We can interpret $\Lambda^2(W)$ as an isotropy representation of parabolic 
subgroup $P$ in a tangent space of ${\cal Q}$ at a point which is fixed by 
$P$. This implies that canonical class $K$ is isomorphic to ${\cal 
O}(-8)$.
\end{proof}

By Borel-Weyl-Bott theorem the algebra $\bigoplus_{n \geq 0} H^0(Q,{\cal 
O}(n))$ is equal to ${\cal S}$. 

It is possible to write a formula for ${\cal S}$ in terms of generators 
and relations. To do this observe that $$\Sym^2(\mathbf{ S})={\cal 
S}_2\oplus \mathbf{ V}$$ Denote $$\Gamma:\mathbf{ V} \rightarrow 
\Sym^2(\mathbf{ S})$$ an inclusion 
of representations . We use the same letter for projection $$\Gamma: 
\Sym^2(\mathbf{ S}) \rightarrow \mathbf{ V}$$ To distinguish these two 
maps we will always specify the arguments. The first map $\Gamma(v)$ has a 
vector argument $v$. The second map $\Gamma(s_1,s_2)$ has two spinor 
arguments $s_1,s_2$.

\begin{proposition}(Cartan)\cite{EC},\cite{Bezr}

{\bf a.} Denote $A_1,\dots,A_{10}$ a basis of $\mathbf{ V}$. Then the algebra ${\cal S}$
defined in (\ref{E:asvdfd}) can be described through generators and relations: $$ {\cal S}= 
\Sym( \mathbf{ S})/(\Gamma(A_1),\dots,\Gamma(A_{10})).$$

{\bf b.} The space $Q$ can be identified with all points $\lambda \in \mathbb{ 
P}(\mathbf{ S}^*)$ such that $\Gamma(\lambda,\lambda)=0.$
\end{proposition}

Consider a complex 
\begin{equation}
Kos^{\bullet}(\mathbf{ S})(i)=\left( 0\rightarrow \Lambda^i(\mathbf{ 
S})\rightarrow \Lambda^{i-1}(\mathbf{ S})\otimes \Sym^1(\mathbf{ 
S})\rightarrow \dots \rightarrow \Sym^{i}(\mathbf{ S})\rightarrow 0 
\right)
\end{equation}
 it is a classical Koszul complex.

The cohomological grading is the degree in the exterior algebra.
The sum $Kos^{\bullet}(\mathbf{ S})=\bigoplus_i Kos^{\bullet}(\mathbf{ 
S})(i)$ is an algebra. It contains $\Sym(\mathbf{ S})=\bigoplus_{i\geq 
0}\Sym^i(\mathbf{ S})$ as a subalgebra. The algebra ${\cal S}$ is a module 
over $\Sym(\mathbf{ S})$. Then
\begin{equation}
B_0= Kos^{\bullet}(\mathbf{ S})\underset{\Sym(\mathbf{ S})}{\otimes} {\cal 
S}=\bigoplus_iQ^{\bullet}(i)
\end{equation}
is called (reduced) Berkovits algebra. The complex $Kos^{\bullet}(\mathbf{ 
S})$ is a free resolution of $\mathbb{ C}$ over $\Sym(\mathbf{ S})$. This 
implies that we have an identity 
\begin{equation}\label{E:vfhop}
H^n(Q^{\bullet}(i))=\Tor_{\Sym(\mathbf{ S})}^{n,i}(\mathbb{ C},{\cal S})
\end{equation}
 (the left upper index corresponds to cohomology index, the right index 
corresponds to homogeneity index). There is a symmetry 
\begin{equation}
\Tor_{\Sym(\mathbf{ S})}^{n,i}(\mathbb{ C},{\cal S})=\Tor_{\Sym(\mathbf{ 
S})}^{n,i}({\cal S},\mathbb{ C}).
\end{equation}
 One way to compute the groups $H^n(Q^{\bullet}(i))$ is to construct a 
minimal resolution of ${\cal S}$ as a module over $\Sym(\mathbf{ S})$ 
(instead of $\mathbb{ C}$ as $\Sym(\mathbf{ S})$ module). Then the 
generators of the modules in the resolution will coincide with cohomology 
classes of the complexes $Q^{\bullet}(i)$. Such resolution was constructed 
in \cite{CR}(though $Spin(\mathbf{ V})$ equivariance in their approach is 
not apparent). Another option is to compute cohomology of $Q^{\bullet}(i)$ 
$i=0,\dots 4$ by brute force using a computer. This has been (partly) done 
in \cite{TS}. In all these approaches due to duality proved in proposition 
(\ref{T:dli}) the only nontrivial task is computation of 
$H^{\bullet}(Q^{\bullet}(4))$.

We are going to construct a (partial) free resolution of $\Sym(\mathbf{ 
S})$ module ${\cal S}$ whose graded components schematically depicted on 
picture (\ref{P:ksnmg})
\begin{equation}\label{P:ksnmg}
\begin{array}{ccccccccccc}
\dots & \dots & \dots & \dots & \dots & \dots & \dots & \dots & \dots & 
\dots & \dots \\
\{0\}& \rightarrow & M_3^3 & \overset{\delta_3}{\rightarrow} & M_3^2 & 
\overset{\delta_2}{\rightarrow} & M_3^1 & \overset{\delta_1}{\rightarrow} 
& M_3^0 & \overset{\delta_0}{\rightarrow} & {\cal S}_3 \\
 & & \{0\} & \rightarrow & M_2^2 & \overset{\delta_2}{\rightarrow} & M_2^1 
& \overset{\delta_1}{\rightarrow} & M_2^0 & 
\overset{\delta_0}{\rightarrow} & {\cal S}_2 \\
 & & & & \{0\} & \rightarrow & M_1^1 & \overset{\delta_1}{\rightarrow} & 
M_1^0 & \overset{\delta_0}{\rightarrow} & {\cal S}_1 \\
 & & & & & & \{0\} & \rightarrow & M_0^0 & \overset{\delta_0}{\rightarrow} 
& {\cal S}_0
\end{array}
\end{equation}
By definition $M^i=\bigoplus_{j\geq 0}M^i_j$ and $M^0=\Sym(\mathbf{ S})$. 

Since algebra ${\cal S}$ is quadratic with ideal of relations 
$I=\bigoplus_{j\geq 1}I_j$ generated by $\mathbf{ V}=I_2$ we have 
$M_2^1=\mathbf{ V},M_1^1=0 $ and $M^1=\mathbf{ V} \otimes \Sym(\mathbf{ 
S})$. Denote $A_i,i=1,\dots,10$ the basis of $\mathbf{ V}$ and 
$u^{\alpha},\alpha=1,\dots,16$ - the basis of $\mathbf{ S}$. The map 
$\delta_1$ is defined by the formula $\delta_1(A_i)=\sum_{\alpha 
\beta=1}^{16}\Gamma_{i\alpha \beta}u^{\alpha}u^{\beta}$. 

The linear space $M^2_3= B$ is the kernel of surjection $\mathbf{ V} 
\otimes \mathbf{ S} \rightarrow I_3$. In the case of interest the 
representation content of $I_3$ is $[1,0,0,1,0]$, representation content 
of $ \mathbf{ V} \otimes \mathbf{ S}_1$ is $[0,0,0,0,1] \oplus 
[1,0,0,1,0]$. It implies that $B=[0,0,0,0,1]=\mathbf{ S}^*$. Denote a 
basis of the vector space $\mathbf{ S}^*$ by $\psi_{\alpha}, 
\alpha=1,\dots,16$. The map $\delta_2$ is given on generators as 
\begin{equation}
\begin{split}
&\delta_2:\mathbf{ S}^*\rightarrow \mathbf{ S}\otimes \mathbf{ V}\\
&\delta_2(\psi_{\beta})=\sum_{\alpha=1}^{16}\sum_{i=1}^{10}\Gamma_{\alpha 
\beta i}A_i u^{\alpha}.
\end{split}
\end{equation}
 We conclude that module $M^2$ is equal to $\mathbf{ S}^*\otimes 
\Sym(\mathbf{ S})+ \tilde{M}^2$, where $\tilde{M}^2$ is a free module with 
generators of degree greater than $3$. We will see later that 
$\tilde{M}^2=0$. Now need to prove a weaker statement: $\tilde{M}^2_4=0$. 
Indeed it is equal to cohomology of the following complex $$\mathbf{ S}^* 
\otimes \mathbf{ S}\overset{\delta_2}{\rightarrow} \mathbf{ V} \otimes 
\Sym^2(\mathbf{ S})\overset{\delta_1}{\rightarrow} I_4\rightarrow 0$$ in 
the term $\mathbf{ V} \otimes \Sym^2(\mathbf{ S})$. We have the following 
representation content:
\begin{equation}
\begin{split}
& I_4=[1,0,0,2,0] \oplus [2,0,0,0,0]\\
& \mathbf{ V} \otimes \Sym^2(\mathbf{ S})=[0,0,0,0,0]\oplus 
[0,0,0,1,1]\oplus [0,1,0,0,0] \oplus [1,0,0,2,0] \oplus [2,0,0,0,0]\\
&\mathbf{ S}^* \otimes \mathbf{ S}=[0,0,0,0,0]\oplus [0,0,0,1,1]\oplus 
[0,1,0,0,0].
\end{split}
\end{equation}
Since the map $\delta_1$ is surjective we need to check that $\mathbf{ 
S}^* \otimes \mathbf{ S}\overset{\delta_2}{\rightarrow} \mathbf{ V} 
\otimes \Sym^2(\mathbf{ S})$ is inclusion. This can be readily checked by 
applying map $\delta_2$ to the highest vectors of each representation in 
decomposition of $\mathbf{ S}^* \otimes \mathbf{ S}$. Let us extend 
partial resolution of $\mathbf{ S}$ to an arbitrary full resolution. Using 
this resolution we can compute $\Tor_{\Sym(\mathbf{ S})}^{i,j}(\mathbb{ 
C},{\cal S})$. Simple computations give the following answer 
\begin{equation}
\begin{split}
&\Tor_{\Sym(\mathbf{ S})}^{0,i}(\mathbb{ C},{\cal S})=\mathbb{ C} \mbox{ 
if } i=0 \mbox{ and } \{0\} \mbox{ if } i\neq 0\\ 
&\Tor_{\Sym(\mathbf{ S})}^{1,i}(\mathbb{ C},{\cal S})=\mathbf{ V} \mbox{ 
if } i=2 \mbox{ and } \{0\} \mbox{ if } i\neq 1\\
&\Tor_{\Sym(\mathbf{ S})}^{2,i}(\mathbb{ C},{\cal S})=\mathbf{ S}^* \mbox{ 
if } i=3 \mbox{ and } \{0\} \mbox{ if } i < 3 \mbox{ or } i=4\\
&\Tor_{\Sym(\mathbf{ S})}^{3,i}(\mathbb{ C},{\cal S})= \{0\} \mbox{ if } 
i<5
\end{split}
\end{equation}

Using the equation (\ref{E:vfhop}) 
and general duality theorem (\ref{T:dli}) and proposition (\ref{T:asfbh}) 
we prove 

\begin{proposition}\label{P:fdt}

The cohomology of the algebra $B_0$ is
\begin{align}\notag
&H^{0,0}=\mathbb{ C}\\ \notag
&H^{1,2}=\mathbf{ V}\\ \notag
&H^{2,3}=\mathbf{ S}^*\\ \notag
&H^{3,5}=\mathbf{ S}\\ \notag
&H^{4,6}=\mathbf{ V}\\ \notag
&H^{5,8}=\mathbb{ C}\\ \notag
&H^{p,q}=0\mbox{ for all $p,q$ not listed above. }\\ \notag
\end{align}%.
\end{proposition}

As we know from \cite{MSch2} the Koszul dual to the algebra ${\cal 
S}=F(\hat{{\cal Q}})$ is the algebra ${\cal S}^!)$ with generators 
$\lambda_{\alpha},\alpha = 1,\dots,16$ which span a linear space $\mathbf{ 
S}^*$ and relations 
\begin{equation}
\sum_{\alpha\beta=1}^{16}\Gamma^{\alpha \beta}_{m_1\dots 
m_5}\{\lambda_{\alpha},\lambda_{\beta}\}=0.
\end{equation} This algebra is a universal enveloping algebra $U(\mathbb{ 
L})$ of a Lie algebra $\mathbb{ L}$ with the same set of generators and 
relations.

The Cartan-Eilenberg complex of a positively graded  Lie algebra $\mathfrak{ g}$ is an 
exterior algebra $\Lambda(\mathfrak{ g}^{\dagger})=\Lambda^{\bullet}(\mathfrak{ g}^{\dagger})$, the dual complex is denoted 
by $\Lambda(\mathfrak{ g})=\Lambda^{\bullet}(\mathfrak{ g})$. The sign $\dagger$ denotes dualisation in a category of graded vector spaces (see Appendix). The 
differential $\Lambda^{\bullet}(\mathfrak{ g}^{\dagger})$ is given by a formula 
\begin{equation}\label{E:kldjws}
(d \nu)(x)=\nu([x_1,x_2])
\end{equation}
where $\nu \in \Lambda^1(\mathfrak{ g}^{\dagger})$ is a linear generator. Denote 
$H^{\bullet}(\mathfrak{ g},\mathbb{ C})$ the homology of such complex, 
homology of the dual complex $\Lambda_{\bullet}(\mathfrak{ g})$ will be denoted as 
$H_{\bullet}(\mathfrak{ g},\mathbb{ C})$ (See \cite{McL} for details about 
(co)homology of Lie algebras). For any positively graded Lie algebra $\mathfrak{ 
g}$ there is a canonical quasiisomorphism $\Lambda_{\bullet}(\mathfrak{ g}) 
\rightarrow \underline{\Barr}(U(\mathfrak{ g})) $ and the dual quasiisomorphism 
$\underline{\Barr}(U(\mathfrak{ g}))^{\dagger}\rightarrow \Lambda(\mathfrak{ g}^{\dagger})$ (see 
\cite{McL} for details).

By one of the properties of Koszul duality transformation $^!$ (see 
\cite{rqwerqe}) there is an inclusion $U(\mathfrak{ g})^!\subset 
H^{\bullet}(\mathfrak{ g},\mathbb{ C})$, for any quadratic Lie algebra.

We need the following  Proposition:
\begin{proposition}\cite{Bezr}
For any compact homogeneous K\"{a}hler manifold $G/P$ of a reductive group 
$G$ and an ample line bundle $\alpha $ on it the Serre algebra 
$\bigoplus_{n \geq 0} H^0(G/P,\alpha^{\otimes n})$ is Koszul.
\end{proposition}
Since the Koszulity relation is reflexive for the case at hand we have an 
isomorphism:
\begin{equation}
U(\mathbb{ L})^!= H^{\bullet}(\mathbb{ L},\mathbb{ C})
\end{equation}
\begin{proposition} 

There is a quasiisomorphism $\rho:\Lambda^{\bullet}(\mathbb{ 
L}^{\dagger})\rightarrow {\cal S}$, which maps linear functional $\lambda^{*\alpha}$ 
on $\mathbb{ L}$ into generator $u^{\alpha}$ of ${\cal S}$. This map is 
zero on subspace $\bigoplus_{i\geq 2}\mathbb{ L}_i^*\subset 
\Lambda^1(\mathbb{ L}^*)$.
\end{proposition}
\begin{proof} 

The only statement which needs to be checked is that $\rho$ commutes with 
differentials. This is obvious, however.
\end{proof}

Our next goal is to relate Berkovits algebra $(B_d,Q)$ with a classical BV 
approach to YM theory. 

Let $\mu:\mathbf{ V}\rightarrow V$ be a linear surjective map. We assume 
that the linear space $\mathbf{ V}$ has an orthonormal basis 
$A_1,\dots,A_{10}$ and linear space $V$ has a basis generated by symbols 
$\frac{\partial}{\partial x^1},\dots,\frac{\partial}{\partial x^d}$. Let 
$\Sym(V^*)$ be a symmetric algebra on the space dual to $V$. The space 
$V^*$ has a basis $x^1,\dots,x^d$. Then $\mu(A_i)$ defines a linear 
functional on $V^*$, which can be extended to a derivation of $\Sym(V^*)$. 
Introduce an algebra 
\begin{equation}
B_{\mu}=B_0\otimes \Sym(V^*)
\end{equation}
and a differential on it by the rule
\begin{equation}
Q=\sum_{\alpha=1}^{16} u^{\alpha}\frac{\partial}{\partial 
\theta^{\alpha}}+\sum_{\alpha 
\beta=1}^{16}\sum_{i=1}^{10}\Gamma_{\alpha\beta}^iu^{\alpha}\theta^{\beta}\mu(A_i).
\end{equation}

Recall that differential algebra $(B_d,Q)$ was defined in the introduction 
in definition (\ref{D:idnsj}) and $(B_{\mu},Q)$ is a minor generalization 
of it.

Consider a graded extension $\mathbb{ M}_{\mu}$ of the Lie algebra 
$\mathbb{ L}$. The linear space $\mathbb{ L}_1=\mathbf{ S}^*$. Then 
$\mathbb{ M}_{\mu 0}=\mathbf{ S}^*$ and this space has a basis 
$\tau_{\alpha},\alpha=1,\dots,16$. A linear space $V$ has a basis 
$\xi_1,\dots,\xi_d$ and $\mathbb{ M}_{\mu 1}=\mathbf{ S}^*+V$. The linear 
space $V$ is by definition is dual to the linear space generated by 
$x^1,\dots,x^d$ of linear coordinates on $d$-dimensional linear space. For 
$i\geq 3$ we have $\mathbb{ M}_{\mu i}=\mathbb{ L}_i$.

The parity of elements of $\mathbb{ M}$ is reduction of grading modulo 
two. The Lie algebra $\mathbb{ M }$ is equipped with a differential 
defined by the formulas:
\begin{equation}
\begin{split}
&d:\mathbb{ M}_{\mu 1}\rightarrow \mathbb{ M}_{\mu 0}; \mathbf{ 
S}^*+V\overset{id,0}{\rightarrow} \mathbf{ S}^*\\
&d:\mathbb{ M}_{\mu 2}\rightarrow \mathbb{ M}_{\mu 1}; \mathbf{ 
V}\overset{0,\mu}{\rightarrow} \mathbf{ S}^*+V\\
&d:\mathbb{ M}_{\mu i}\rightarrow \mathbb{ M}_{\mu i-1}; i\geq 3, d=0
\end{split}
\end{equation}
The commutation relations in the algebra $\mathbb{ M}$ are of semidirect 
product $\mathbb{ L}\ltimes (\mathbf{ S}^*+V)$, where $\mathbf{ 
S}^*\subset \mathbb{ M}_{\mu 0}, V\subset \mathbb{ M}_{\mu 1}$. The linear 
space $ \mathbf{ S}^*+V $ is an abelian ideal. The action of $\mathbb{ L}$ 
on $\mathbf{ S}^*+V$ is given by the rule:
\begin{equation}
\begin{split}
&[\theta_{\alpha},\tau_{\beta}]=2 \mu \sum_{i=1}^{10} \Gamma^i_{\alpha 
\beta}A_i\\
&[\theta_{\alpha},\xi_i]=0
\end{split}
\end{equation} 
One can consider a version of Cartan-Eilenberg complex for differential 
graded Lie algebra $(\mathbb{ M}_{\mu},d)$, in the complex 
$\Lambda^{\bullet}(\mathbb{ M}_{\mu}^{\dagger})$ the formula (\ref{E:kldjws}) 
becomes
\begin{equation}
(D \nu)(x)=\nu([x_1,x_2])+\nu(d(x_3)).
\end{equation} 

There is a map 
\begin{equation}
\chi:(\Lambda(\mathbb{ M}_{\mu}^{\dagger}),D)\rightarrow (B_{\mu},Q).
\end{equation}
On the generators the map is 
\begin{equation}
\begin{split}
&\chi(\xi^{*i})=x^i\\
&\chi(\lambda^{*\alpha})=u^{\alpha}\\
&\chi(\tau^{\alpha})=\theta^{\alpha}\\
&\chi\ \mbox{ is zero on the rest of the generators. }
\end{split}
\end{equation}
To make this map a map of complexes one has to modify slightly the grading 
on $B_{\mu}$:
\begin{equation}
\begin{split}
&\tilde{x^i}=2\\
&\tilde{u^{\alpha}}=2\\
&\tilde{\theta^{\alpha}}=1
\end{split}
\end{equation}
The grading on $\Lambda(\mathbb{ M}^{\dagger}_{\mu})$ is a standard cohomological 
grading.
\begin{proposition}\label{P:anbx}
The map $\chi$ is correctly defined and is a quasiisomorphism of algebras $(\Lambda(\mathbb{ M}_{\mu}^{\dagger}),D)$ and $(B_{\mu},Q)$.
\end{proposition}
\begin{proof} 

We leave the proof of the first statement as an exercise for the reader.
The algebra $\mathbb{ M}_{\mu}$ carries an action of $\mathbb{ 
C}^{\times}$ which commutes with differential $D$. It manifests itself in 
a grading. In this grading $DEG(\lambda_{\alpha})=1$. This condition 
allows uniquely spread the grading on the entire algebra. The induced 
grading on $\Lambda(\mathbb{ M}_{\mu})$ has the following feature: all 
graded components become finite-dimensional bounded from both sides 
complexes . Such grading could be pushed on $B_{\mu}$. A simple 
observation is that the map $\chi$ is surjective.
A filtration of $\Lambda(\mathbb{ M}_{\mu})$ which leads to 
Serre-Hochschild spectral sequence, based on the extension 
\begin{equation}
0\rightarrow \mathbf{ S}^*+V\rightarrow \mathbb{ M}_{\mu}\rightarrow 
\mathbb{ L}\rightarrow 0
\end{equation}
 can be pushed to the algebra $B_{\mu}$. The $E_2$-terms of the 
corresponding spectral sequences are isomorphic to algebra $B_{\mu}$. 
Therefore the limiting terms of the spectral sequnce, which converge 
strongly, must coincide.
\end{proof}

\begin{proposition}\label{P:iwjshdf}
The universal enveloping algebra $U(\mathbb{ M}_{\mu})$ is Koszul dual to 
$B_{\mu}$.
\end{proposition}
\begin{proof} 

The proof is a 
straightforward application of definitions.
\end{proof}

In order to avoid confusion, when we talk about differential Lie algebras 
by (co)homology we always mean cohomology of Cartan-Eilenberg complex. 
However the linear space of the algebra itself carries a differential 
which we call intrinsic differential. Cohomology of such differential will 
be called intrinsic cohomology.

The algebra $\mathbb{ M}_{\mu}$ carries an intrinsic differential. It 
allows to reduce the space of the algebra without affecting its 
cohomology. Introduce two subalgebras $E_{\mu}\mathbb{ M}=\bigoplus_{i\geq 
1} E_{\mu}\mathbb{ M}_i$ and $T_{\mu}\mathbb{ M}=\bigoplus_{i\geq 
2}T_{\mu}\mathbb{ M}$ of $\mathbb{ M}_{\mu}$:
\begin{equation}
\begin{split}
&E_{\mu}\mathbb{ M}_1=V\\
&E_{\mu}\mathbb{ M}_i=\mathbb{ L}_i,i\geq2\\
&\mbox{ the differential is a restriction of the differential on } 
E_{\mu}\mathbb{ M}
\end{split}
\end{equation}

\begin{equation}
\begin{split}
&T_{\mu}\mathbb{ M}_2=Ker[\mu :\mathbb{ L}_2\rightarrow V]\\
&T_{\mu}\mathbb{ M}_i=\mathbb{ L}_i,i\geq3\\
&d=0
\end{split}
\end{equation}
\begin{proposition}\label{P:ksksj}
The algebras $E_{\mu}\mathbb{ M}$, $T_{\mu}\mathbb{ M}$ 
quasiisomorphically embed into algebra $\mathbb{ M}_{\mu}$.
\end{proposition}
\begin{proof} 

Obvious.
\end{proof}

\begin{corollary}
$H^{\bullet}(T_{\mu}\mathbb{ M},\mathbb{ C})=H^{\bullet}(E_{\mu}\mathbb{ 
M},\mathbb{ C})=H^{\bullet}(B_{\mu})$, where the first two groups are cohomology of Lie algebras. The last group is intrinsic of differential algebra $B_{\mu}$.
\end{corollary}

We need to identify algebras $T_{\mu}\mathbb{ M}$ and $E_{\mu}\mathbb{ 
M}$. Assume that $\mu=0$. In \cite{MSch2} we indicated that the algebra 
$\mathbb{ L}$ contains a homomorphic image of Lie algebra $\mathbb{S}YM$. 
The universal enveloping algebra $U(\mathbb{ S}YM)$ is isomorphic to 
$YM_0$ with $D=10,d'=0,N=1$ and potential $U$ equal to zero, The Lie 
algebra $\mathbb{S}YM$ is generated by 
\begin{equation}\label{E:jkrhd}
\begin{split}
&A_i=\sum_{\alpha \beta}\Gamma_{i}^{\alpha 
\beta}\{\lambda_{\alpha},\lambda_{\beta}\}\\
&\psi^{\alpha}=\sum_{\beta}\sum_{m=1}^{10}\Gamma^{\alpha \beta 
m}[\lambda_{\beta},A_m]
\end{split}
\end{equation}
with relations (\ref{E:relssfs1},\ref{E:relssfs2},\ref{E:relssfs3}).
\begin{proposition}\label{P:jkdhjas}
The Lie algebra $\mathbb{S}YM$ is isomorphic to the algebra 
$\bigoplus_{i\geq 2}\mathbb{ L}_i$.
\end{proposition}
\begin{proof} 

By construction $\mathbb{S}YM$ maps in $\bigoplus_{i\geq 2}\mathbb{ 
L}_i$. We need to check that the set of generators of $\bigoplus_{i\geq 
2}\mathbb{ L}_i$ coinsides with 
$A_1,\dots,A_{10},\psi^{1},\dots,\psi^{16}$ and there is no relations 
other then (\ref{E:relssfs1},\ref{E:relssfs2},\ref{E:relssfs3}) with 
$D=10,d'=0,N=1$ and $U=0$. To do so we take advantage of the proposition 
(\ref{P:h12}) and corollary (\ref{C:kgdfgh}). The space 
$H^{\bullet}(E_0\mathbb{ M},\mathbb{ C})$ is quasiisomorphic to 
$H^{\bullet}(B_0)$ which were computed in proposition (\ref{P:fdt}). 
According to this proposition $H^{1}(E_0\mathbb{ M},\mathbb{ C})=\mathbf{ 
V}+\mathbf{ S}^*$ and $H^{2}(E_0\mathbb{ M},\mathbb{ C})=\mathbf{ 
V}+\mathbf{ S}$. It implies th that elements $A_1,\dots,A_{10}$ which span 
$\mathbf{ V}$ and $\psi^{1},\dots,\psi^{16}$ which span $\mathbf{ S}$ are 
indeed the generators of the algebra $E_0\mathbb{ M}$. The relations 
(\ref{E:relssfs1}) (if we think of them as of elements of a free algebra) 
transform as representation $\Pi \mathbf{ V}^*$, (\ref{E:relssfs1}) 
transform as representation $\mathbf{ S}^*$ are indeed the generators of 
the ideal of relations of the algebra $E_0\mathbb{ M}$.
\end{proof}

If $\mu=0$ then $\mathbb{ M}_{\mu}$ is quasiisomorphic to $T_0\mathbb{ 
M}=\bigoplus_{i\geq 2} \mathbb{ L}_i$.
\begin{proposition}\label{E:jkldfhja}
The Koszul dual to $B_0$ is quasiisomorphic to $U(\mathbb{ S}YM)$.
\end{proposition}
\begin{proof} 

According to proposition (\ref{P:jkdhjas}) $U(\mathbb{ S}YM)$ is 
quasiisomorphic to $U(\bigoplus_{i\geq 2} \mathbb{ L}_i)$. By the remark 
from the previous paragraph $\bigoplus_{i\geq 2} \mathbb{ L}_i$ is 
quasiisomorphic to $T_0\mathbb{ M}$ which is by proposition 
(\ref{P:ksksj}) is quasiisomorphic to $\mathbb{ M}_0$. By proposition 
(\ref{P:iwjshdf}) $\mathbb{ M}_0$ is Koszul dual to $B_0$.
\end{proof}

\begin{corollary}\label{C:ogsw}
There is a quasiisomorphism $bv_0^*\cong B_0$.
\end{corollary}
\begin{proof} 

We already established a quasiisomorphism $\Lambda(E_0\mathbb{ M})$ and 
$B_0$. We have a canonical quasiisomorphism $\underline{\Barr}(E_0\mathbb{ 
M})^{\dagger}\rightarrow \Lambda(E_0\mathbb{ M}^{\dagger})$. On the other hand according to 
proposition (\ref{T:fff}) we have a quasiisomorphism $\Barr YM \cong 
bv_0$. Dualisation of the last quasiisomorphism gives the necessary 
quasiisomorphism.
\end{proof}

\begin{proposition}\label{P:kdjdgb}
There is a quasiisomorphism
\begin{equation}\label{E:lskdhbrt}
B_{\mu}\cong bv_{\mu}^*.
\end{equation}%.
\end{proposition}
\begin{proof} 

There is an obvious identification $U(T_{\mu}\mathbb{ M})\cong T_{\mu}YM$ 
($U$ stands for universal enveloping algebra) which comes from 
identification $T_{0}\mathbb{ M}\cong \mathbb{ S}YM$. Then by lemma 
(\ref{L:lknb}) and propositions (\ref{E:hrtfds}) and (\ref{P:anbx}) we 
have a quasiisomorphism (\ref{E:lskdhbrt}).
\end{proof}
\begin{proposition}\label{P:qraffcsd}
Koszul dual to $B_{\mu}$ is quasiisomorphic to $U(T_{\mu}\mathbb{ M})$.
\end{proposition}
\begin{proof} 

According to proposition (\ref{P:iwjshdf}) the Koszul dual to $B_{\mu}$ is 
equal to $U(\mathbb{ M}_{\mu})$. The result follows from the previous 
observation and proposition (\ref{P:ksksj}).
\end{proof}

\section{Appendix \\Dualisation  in the category of linear spaces \\Bar duality}

\subsection{Dual spaces}\label{S:dual}
Suppose we have an inverse system of finite dimensional vector spaces 
$\rightarrow N^{i+1}\overset{i}{\rightarrow} N^{i}\rightarrow \dots 
\rightarrow N^{0}$ ($i\geq 0$), where all maps $i$ are surjective. Let  
$N=\Llim{i}N^i$. There is a canonical map $N\rightarrow N^{i}$ which in our 
case is surjective. Denote the kernel if this map by $J^i$. It is clear 
that $J^{i-1}\subset J^{i}$ and the set of linear spaces $J^{i}$ 
completely determines the inverse system, and for every linear space $W$ 
with an decreasing filtration $J^i$ such that 
\begin{align}
& \bigcap_{\infty}^{i=0} J^i=\{0\}\label{E:decr}\\
& dim(J^i/J^{i-1})< \infty \label{E:fin}
\end{align}
we have $W \subset \Llim{i}{W/L_i}=\widehat{W}$. 
Define a direct system of finite dimensional vector spaces $M^{0} 
\rightarrow M^{-1} \rightarrow \dots M^{n} \rightarrow M^{n-1} \rightarrow 
\dots $ as $M^{n}=N^{*-n}$. We call such direct system dual to $N^n$ and 
denote it by $N^{*n}$. It should be clear how to define $N^{**n}$ and that 
it is equal to $N^{n}$. Observe that $\colim{n}N^{*n}$ has an increasing 
filtration by spaces $F^i=N^{i*}$. Denote $M=\colim{n} N^{*n}$. The 
filtration satisfies 
\begin{align}
&M=\bigcup_i F^i \label{E:comp} \\
& dim(F^{i+1}/F^i)< \infty . \label{E:fin1}
\end{align}

Skipping all mentionings of limits we can say that there is a dualization 
invertible functor from category of complete linear spaces with decreasing 
filtration $W,J_i\quad (i\leq 0)$, where $J_i$ satisfies (\ref{E:decr}, 
\ref{E:fin}) and linear spaces equipped with increasing filtration $M,F^i$ 
such that $F^i$ satisfies (\ref{E:comp}) and (\ref{E:fin}). We will refer 
to such duality as {\bf topological}.
\begin{definition}
Denote $U=\bigoplus_{i \in \mathbb{ Z}} U_i$ a graded vector space with 
$dim(U_i) <\infty$. One can define a dualization functor on the category 
of such vector spaces. Indeed by definition $U^{\dagger}=\bigoplus_{i \in 
\mathbb{ Z}} U^*_{-i}$, the grading of $U^*_i$ is equal to $-i$. Observe 
that the functor $U^{\dagger}$ is auto duality in the category of graded 
linear spaces. A vector space in this since to $U$ will be called {\bf 
algebraic } dual.
\end{definition}

Given a graded vector space $U=\bigoplus_{i \geq 0} U_i$ one can define 
two linear spaces with filtrations: 

{\bf a.} $U=\bigoplus_{i \geq 0} U_i$ and a filtration is defined as 
$F^n=\bigoplus_{0\leq i \leq n} U_i$.

{\bf b.} $\widehat{U}=\prod_{i \geq 0} U_i$ and a filtration is defined as 
$J_n=\prod_{n\leq i } U_i$. In future the sign $\widehat{}$ will always 
means completion of the space $W$ with respect to a decreasing filtration 
.

In future if we write $U^*$ for $U=\bigoplus_{i \geq 0} U_i$ we will 
always mean the topological dual.

\subsection{$A_{\infty}$-algebras}\label{S:mmcsa}

  Let us consider a $\mathbb{Z}_2$-graded  vector space $W$ and corresponding tensor algebra $T(W)=\bigoplus_{n\geq 1}W^{\otimes n}$. The tensor algebra $T(W)$ is  $\mathbb{Z}$-graded, but it has also   $\mathbb{Z}_2$-grading  coming from   $\mathbb{Z}_2$-grading of $W$. We say that differential (= an odd derivation having zero square) $Q$ on   $T(W)$  specifies a structure of  $A_{\infty}$-coalgebra on $V=\Pi W$. 

One can describe the structure of   $A_{\infty}$-coalgebra on  $V=\Pi W$ as a sequence of linear maps $\Delta_1: V\rightarrow V,\ \  \Delta_2:V \rightarrow V \otimes V,\ \  \Delta_n:V\rightarrow V^{\otimes n} $. Using the Leibniz rule we can extend $\Delta_1,\Delta_2\dots$ to a derivation $Q$ of  $(T(W)$;       the condition $Q^2=0$ implies some conditions on $\Delta_1, \Delta_2,\dots$                   
The map $\Delta_1$ is a differential $(\Delta_1^2=0)$,  the map $\Delta_2$ can be interpreted as comultiplication. If $\Delta_n=0$ for $n\geq 3$ we obtain a structure of associative coalgebra on $V$.

One says that differential algebra  $(T(W),Q)$ is (bar-) dual to  $A_{\infty}$-coalgebra $(V,m)$ or that  $(T(W),Q)$ is obtained  from $(V,\Delta)$ by means of bar-construction. We will use the notation $\Barr(V, \Delta)$ for this differential algebra. Hochschild  homology of $(V,\Delta)$ are defined as homology of  $(T(W),Q)$.

An   $A_{\infty}$-map of coalgebras is defined as a homomorphism of corresponding differential tensor algebras (i.e. as an even homomorphism  $\Barr(V, \Delta) \rightarrow \Barr(V^{\prime}, \Delta ^{\prime})$ commuting with differential).

One can describe an $A_{\infty}$-map $\phi(V,\Delta)\rightarrow (V^{'},\Delta^{'})$ by means of a sequence of maps $\varphi _{n}V\rightarrow V^{' \otimes n}$ where $V=\Pi W,V^{'}=\Pi W^{'}$. 

The map $\varphi_{1}$ commutes with differentials $(\Delta^{'}_{1}\varphi_{1}=\varphi_{1}\Delta_{1})$, hence it induces a homomorphism of homology of $(V,\Delta_{1})$ into homology of $(V^{'},\Delta_{1}^{'})$.  If the induced  homomorphism is an isomorphism one says that $A_{\infty}$-map is a quasiisomorphism.

One can define an $A_{\infty}$-algebra structure on $\mathbb{Z}_2$-graded vector space $V$ as an odd coderivation $Q$ of tensor coalgebra $T(W)=\bigoplus _{n\geq 1}W^{\otimes n}$ obeying $Q^{2}=0$. (Here $W=\Pi V$). 

Equivalently , $A_{\infty}$ algebra $(V,m)$ can be defined as $\mathbb{Z}_2$-graded vector space $V$ equipped with a series of operations 

$$m_{1}:V\rightarrow V,m_{2}:V^{\otimes 2}\rightarrow V,\dots ,m_{n}:V^{\otimes n}\rightarrow V$$, obeying certain conditions.
 
One says that the differential coalgebra $(T(W),Q)$ is bar-dual to $A_{\infty}$ algebra $(V,m)$ or that $(T(W),Q)$ is obtained from $(V,m)$ by means of bar-construction. We will use the notation $\Barr(V,m)$ for differential coalgebra $(T(W),Q)$. Hochshild homology of $(V,m)$ is defined as homology of $(T(W),Q)$. 

Usually one considers bar-duality for 
$\mathbb{Z}$-graded $A_{\infty}$-algebras 
and $A_{\infty}$-coalgebras.

The algebra 
$\Barr (V,\Delta)=(T(W),Q) $ 
that is dual to $\mathbb{Z}$-graded$A_{\infty}$-coalgebra $(V,\Delta)$, will 
be considered as $\mathbb{Z}$-graded differential algebra. 
The space $W$ is equal $V[-1]$; in other words 
$W$ coincides with $V$ with grading shifted 
by $-1$. Similarly, 
the $\mathbb{Z}$-graded coalgebra $\Barr(V,m)$ dual to graded 
$A_{\infty}$-algebra $(V,m)$ will be considered as 
graded differential coalgebra.

Let us consider the case of  $A_{\infty}$-(co)algebras $(V,m)$ and $(V^{'},m)$ have an additional  positive grading. This is an auxiliary grading which has no correlation with internal 
(homological) $\mathbb{Z}_2(\mathbb{Z})$-grading. We assume that all structure maps $m_k$ have degree zero with respect do this additional grading. The same applies to A$_{\infty}$-morphisms. Such $A_{\infty}$-(co)algebras will be called homogeneous. We will use an abbreviation h.morphism h.quasiisomorphism etc. for homogeneous morphism, homogeneous quasiisomorphism ...

\begin{theorem} Two homogeneous $\Barr(V,\Delta)$ 
s 
$A_{\infty}$-(co)algebras are h.quasiisomorphic iff their dual algebras (coalgebras) are h.quasiisomorphic. 
\end{theorem}

\begin{theorem}
 If $A_{\infty}$-morphism $f:(V,m)\rightarrow (V^{'},m^{'})$ of homogeneous (co)algebras induces an isomorphism of Hochshild homology then $f$ is h.quasiisomorphism.
\end{theorem}

%A morphism of $A_{\infty}$-(co)algebras is graded if its components have zero degree with respect to additional grading.
%Using the fact, that by definition Hochshild homology  $A_{\infty}$-coalgebra is a homology of dual differential algebra, the following immediate consequences of this statement. 
%
%\begin{theorem} A graded $A_{\infty}$-map of graded $A_{\infty}$-coalgebras is a quasiisomorphism iff it is quasiisomorphism of  Hochshild homology of this algebras. 
%% Using the fact, that by definition Hochshild homology ?? $A_{\infty}$-coalgebra is a homology of dual differential algebra.
%\end{theorem}
For quadratic algebras bar-duality is closely related to Koszul duality. Let $A$ be a quadratic algebra $A=\bigoplus_{n>0} A^{n}, dim A^{n}<\infty $ and $B=\bigoplus_{n>0} A^{*n}$ be the dual graded coalgebra. 

\begin{proposition} 

The differential graded algebra bar-dual to the coalgebra $B$ is quasiisomorphic to the Koszul dual $A^{!}$ if $A$ is a Koszul algebra.
\end{proposition}

We will consider duality in more general situation when $A_{\infty}$-coalgebra $(V,\Delta)$ whose descending filtration $F^{k}$ obeying $F^{1}=V,\cap_{k\geq 1} F^{k}=0$ and $V$ is complete with respect to filtration.
Then we can introduce corresponding filtration $F^{k}$ on $T(W)$.
 We define $F^{p}(T(W))$ by the formula 
\begin{equation}\label{E:fnsllgg2}
\sum_{\sum_{r=1}^k n_r\geq p} F^{n_1}\otimes \dots \otimes F^{n_{k}}.
\end{equation}
 We assume that the structure 
of $A_{\infty}$-coalgebra is compatible with filtration. This means that 
\begin{equation}\label{jadytvb}
\Delta_k(F^s)\subset \sum_{n_1+ \dots +n_k\geq s} F^{n_1}\otimes \dots 
\otimes F^{n_k},\quad n_k\geq 1.
\end{equation}

 In the language of tensor algebras we require that $Q(F^{k}(T(W)))\subset F^{k}(T(W))$.
 In particular, for filtered $A_{\infty}$-coalgebra we have $\Delta _{1}(F^{s})\subset F^{s}$ hence we can consider homology of $(F^{s},\Delta _{1})$.

The bar dual to the filtered $A_{\infty}$-coalgebra $(V,\Delta)$ is defined as topological differential algebra $(\widehat{T(W)},Q)$ obtained from $(T(W),Q)$ by means of completion with respect to filtration $F^{k}$.

$A_{\infty}$-maps of filtered coalgebra should agree with filtrations; they can by considered as continuous homomorphisms of dual topological differential algebras.

Representing $A_{\infty}$-map $\phi:(V,\Delta _{k})\rightarrow (V^{'},\Delta ^{'}_{k})$ as a series of maps $\varphi_{k}:V\rightarrow V^{'\otimes k}$ and using that $\Delta ^{'}_{1}\varphi_{1}=\varphi_{1}\Delta_{1}\quad \phi_{1}$ 
induces a homomorphism of homology of $(F^{k}/F^{k+1},\Delta_{1})$ into homology of $(F^{' k}/F^{' k+1},\Delta^{'}_{1})$, if all of these homomorphisms are isomorphisms we say that $\phi$ is a filtered quasiisomorphism. There are notfiltered quasiisomorphisms between filtered objects.

We introduce also a notion of filtered $A_{\infty}$-algebra $(V,m)$ fixing a decreasing filtration $F^{p}$ on $V$ $p\geq 1$that satisfies the following conditions: 
 \begin{equation}\label{E:jkuisa}
\mu_k:F^{s_1}\otimes \dots \otimes F^{s_k}\rightarrow F^{s_1+\dots s_k}. 
\quad k \geq 1
\end{equation}\
\begin{equation}
\bigcap_sF^{s}=0\quad F^1=V
\end{equation}%
and  $V$ is complete with respect to such filtration.
 (Notice that the notion of filtered $A_{\infty}$-algebra is not dual to the notion of filtered $A_{\infty}$-coalgebra ( a filtration that is dual to decreasing filtration is an increasing filtration). 
The differential coalgebra $\Barr(V,m)$ corresponding to filtered $A_{\infty}$-algebra can be considered as filtered coalgebra (see formula (\ref{jadytvb}) for filtration). Its completion $\wBar(V,m)$ also can be regarded as a filtered differential topological coalgebra. 

Let $f$ be an $A_{\infty}$-morphism of filtered $A_{\infty}$-algebras $(V,m)\rightarrow (V^{'},m^{'})$ that is compatible with filtrations.
It induces a map $f_{*}:\Barr(V,m)\rightarrow \Barr(V^{'},m^{'})$ of corresponding dual coalgebras, that can be extended to a map $\hat{f_{*}}:\widehat{\Barr(V,m)}\rightarrow \widehat{\Barr(V^{'},m^{'})}$.

We need the following statements proved in \cite{M1}:
\begin{theorem} 

If filtered $A_{\infty}$-coalgebras are quasiisomorphic then dual topological algebras are quasiisomorphic.
\end{theorem}
\begin{theorem}\label{T:lskdfh}
Let $(V,\Delta)$ and $(V^{'},\Delta ^{'})$ be two filtered $A_{\infty}$-algebras . Then quasiisomorphism of corresponding topological differential coalgebras $(\widehat{T(W)},Q)$ and $(\widehat{T(W^{'})},Q^{'})$ implies quasiisomorphism of $A_{\infty}$-algebras $(V,\Delta)$ and $(V^{'},\Delta ^{'})$.

Let $(V,\Delta)$ and $(V^{'},\Delta ^{'})$ be two filtered $A_{\infty}$-algebras . Then a filtered quasiisomorphism of $(V,\Delta)$ and $(V^{'},\Delta ^{'})$ implies quasiisomorphism of corresponding topological differential coalgebras $(\widehat{T(W)},Q)$ and $(\widehat{T(W^{'})},Q^{'})$.
\end{theorem}

\begin{theorem} 

Let $(V,\Delta)$ and $(V^{'},\Delta ^{'})$ be two filtered $A_{\infty}$-coalgebras . Then filtered quasiisomorphism of corresponding topological differential algebras $(\widehat{T(W)},Q)$ and $(\widehat{T(W^{'})},Q^{'})$ implies quasiisomorphism of $A_{\infty}$-coalgebras $(V,\Delta)$ and $(V^{'},\Delta ^{'})$.
\end{theorem}
\begin{theorem}

If the map $f_*:\wBar(V,m)\rightarrow \wBar(V^{'},m)$ is a quasiisomorphism that the original map $f$ is also a quasiisomorphism.
\end{theorem}
\begin{lemma} \label{L;sssfg}
For any A$_{\infty}$ filtered coalgebra $H$ there is A$_{\infty}$-morphism
\begin{equation}\label{E:fghb}
\wBar \wBar(H)\overset{\psi}{ \rightarrow} H
\end{equation} of A$_{\infty}$-coalgebras.
The morphism $\psi$ is a quasiisomorphism.

Similarly for any A$_{\infty}$ filtered algebra $A$ there is A$_{\infty}$i-morphism
\begin{equation}\label{E:fghb1}
A\overset{\phi}{ \rightarrow} \wBar \wBar(A)
\end{equation} of A$_{\infty}$-coalgebras.
The morphism $\phi$ is a quasiisomorphism.
\end{lemma}
\subsection*{Acknowledgments}
We would like to thank  A. Bondal, A. Gorodenzev, M. Kontsevich,  D. Piontkovsky, A. Rudakov, E. Witten for
stimulating discussions. Part of this work was done when one or both 
of the authors were staying in  
Mittag-Leffler Institute; we  appreciates the 
hospitality of this institution.

\end{document}